\documentclass{article}
\usepackage{amsmath}
\usepackage{amsfonts}
\usepackage{amssymb}

\begin{document}
\bigskip
%\pagecolor{black}
%\color{white}
%
%
%
%
\begin{titlepage} \vspace{0.2in}
\begin{center} {\LARGE\bf
A New Approach in Quantum Gravity and its Cosmological Implications\\}%
\vspace*{0.8cm}
{\bf Simone Mercuri}\\ \vspace*{0.3cm}
{\bf Giovanni Montani}\\ \vspace*{1cm}
ICRA---International Center for Relativistic Astrophysics\\
Dipartimento di Fisica (G9),\\
Universit\`a  di Roma, ``La Sapienza",\\
Piazzale Aldo Moro 5, 00185 Rome, Italy.\\
e-mail: mercuri@icra.it, montani@icra.it\\
\vspace*{1.8cm}
\vspace*{1cm} \\
{\bf Abstract  \\ } \end{center} \indent
This work concerns a new reformulation of quantum geometrodynamics, which allows to overcome a fundamental ambiguity contained in the canonical approach to quantum gravity: the possibility of performing a (3+1)-slicing of space-time, when the metric tensor is in a quantum regime. Our formulation provides also a procedure to solve the problems connected to the so called \emph
{frozen formalism}. In particular we fix the reference frame (i.e. the lapse function and the shift vector) by introducing the so called \emph{kinematical action}; as a consequence, the new hamiltonian constraints become parabolic, so arriving to evolutive (Schr\"odinger-like) equations for the quantum dynamics.\newline
The kinematical action can be interpreted as the action of a pressure less, but, in general, non geodesic perfect fluid, so in the semi classical limit our theory leads to the dynamics of the gravitational field coupled to a dust which represents the material reference frame we have introduced fixing the slicing. We also investigate the cosmological implications of the presence of the dust, which, in the WKB limit of a cosmological problem, makes account for a \emph{dark matter} component and could play, at present time, a dynamical role.
\end{titlepage}

\section{Introduction}

\label{intro}

\bigskip

Time has a special role in all the classical \footnote{With the adjective
``classical'' we intend ``not general relativistic''.} theories of physics.
Newton's time is an external parameter, respect to which we describe the
dynamics of the system. In non relativistic quantum mechanics time is not a
physical observable in the usual sense, i.e. it does not exist an operator
associated to the time variable, but it is an external parameter as well as in
classical mechanics. The construction of the theory is deeply influenced by
the concept of an external time, for example, to construct the Hilbert space
of quantum states we have to choose a complete set of observables, which
commute at equal instants of time. It follows that the dynamical equation for
the non relativistic quantum mechanics has an explicit time dependence, which
reflects on the evolutive character of the quantum states represented by wave
functionals. Moreover the special role of time is also the reason why the
time-energy indetermination,
\begin{equation}
\Delta t\Delta E\geq\hbar,
\end{equation}
has a different meaning respect to the one usually associated to position and momentum.

All these simple ideas can be generalized to those systems compatible with the
special theory of relativity. In this case Newton's time is replaced by the
time measured in a set of relativistic inertial frames, but the space-time is
an external non dynamical structure yet, profoundly different with respect to
the dynamical one of general relativity.

To construct a consistent quantum field theory (in the canonical approach) we
need the hamiltonian function, which is the conjugate momentum to the
relativistic temporal coordinate. So, in order to apply the canonical
quantization procedure to the gravitational field, we must, first of all,
extract a possible time parameter from the classical theory; but this is not a
simple task for a diffeomorphisms invariant theory like general relativity.
Moreover the canonical quantization procedure leads, as well known, to the so
called Wheeler-De Witt approach \cite{DeW1967, DeW1997}, which does not contain
any evolution with respect to the time parameter, reflecting the invariance
under diffeomorphisms of the classical theory, i.e. the lack of any external
temporal parameter in general relativity.

In this work we introduce the canonical quantization program,
dedicating a wide review either to the Arnowitt-Deser-Misner formalism (ADM)
\cite{ArnDesMis1959, ArnDesMis1960a, ArnDesMis1960b, ArnDesMis1962}, either to
the Wheeler-DeWitt equation (WDE). \newline ADM formalism is the way to
extract a ``time'' dependence from the gravitational theory. It is based on
the (3+1)-\emph{slicing} of the space-time, where ``time'' plays the role of
parameter for the foliation, singling out the different elements of a
particular family of hypersurfaces, which fill the whole space-time. It is
worth noting that, in the classical theory, the slicing procedure is well
defined and gives a time dependence to the events by spacial frames
represented by the hypersurfaces, but, in the quantum formulation of the
theory it generates ambiguities; in fact, when the metric tensor is in a
quantum regime, defining the space or time like character of a vector field
(necessary to develop the slicing procedure) becomes an impossible task:
it seems possible to distinguish between space or time like vector field only
in average (expectation values) sense. \newline It is just in these
ambiguities that our criticism to the canonical quantization of gravity
arises. We claim that to give sense to the slicing procedure also in a quantum
regime it is necessary to fix a reference frame with respect to which to
operate the slicing \cite{Mon2002, MerMon2003}.

Another important feature of general relativity for this discussion is the so
called \emph{relationalism}. We know that the diffeomorphisms invariance of
the classical theory requires the absence of any non dynamical object in the
theory, in particular, in general relativity, the space-time itself is a
dynamical field. So differently from the Newtonian mechanics we have not a
fixed background on which we can localize the physical events; but the
localization is fully relational, in other words a dynamical object can be
localized only with respect to another one. These considerations lead to think
that a reference frame in the gravitational theory has to be a dynamical
physical entity coupled to the gravitational field. \newline There exist two
different way to introduce a reference system in a classical or quantum
gravity theory, the first one consists in adding to the gravitational field a
dynamical fluid or fields \cite{Rov1991a, Rov1991b} and \cite{Smo1993}, the
other one in fixing the frame in geometrical way \cite{KucTor1991, BicKuc1995}
(see also \cite{Kuc1991, Kuc1992, Kuc1992 2}), i.e. fixing the splitting. We
stress that in a recent paper is shown that these two approaches lead to an
equivalent evolutive quantum dynamics, in other words there exists a dualism
between introducing a physical frame and breaking down time diffeomorphisms
invariance \cite{MerMon2003b}. It is worth noting that all these approaches
lead to a Schr\"{o}dinger-Einstein quantum dynamics, i.e. the introduction of
a material reference frame in the classical dynamics leads to an evolutive
quantum equation for the dynamics of the coupled system.

Our point of view is the geometrical one: we fix the reference frame choosing
a particular family of hypersurfaces, assigning particular values to the lapse
function and to the shift vector \cite{Mon2002, MerMon2003}. It is clear that
in this way we loose the hamiltonian constraints, and so the possibility to
canonically quantize the system. But using the so called \emph{kinematical
action}, already introduced in the quantum field theory on curved background
\cite{Kuc1981} to reparametrize the gravitational action, we obtain new
hamiltonian constraints; so the introduction of the kinematical action is the
price we have to pay to perform the canonical quantization after having fixed
the slicing.

The kinematical action is a geometrical object, which links the choice of the
lapse function and the shift vector to a particular family of hypersurfaces
and to their normal vector field, but it has also a clear physical
interpretation. In fact, in section \ref{sec2 ka}, it is possible to show that the
kinematical term is, the action of a non relativistic dust, which
couples with a gravito-electromagnetic-like field. \newline An important
outcome of our theory is the appearance in the new hamiltonian constraints of
a linear term, strictly connected to the kinematical action, which gives an
evolutive character to the quantum equation, i.e. the equations become
parabolic \cite{Mon2002, MerMon2003}. This feature is not so unexpected,
because, fixing the lapse function and the shift vector, we are breaking the
diffeomorphisms invariance of the theory, so arriving to evolutive equations
along the fixed slicing.

The physical interpretation of the kinematical action allows us to recognize
in the temporal parameter the conjugate variable to the energy density of the
dust. The evolutive theory, moreover, overcomes the well known shortcomings of
the WDE approach as shown in this paper as well as in \cite{Mon2002,
MerMon2003}. Moreover to study the phenomenology connected with the appearance
of this additional energy term we apply our theory to a cosmological model. In
particular we make some estimations to understand if this new energy term has
something to do with the observed \emph{dark matter} of the Universe.

Section \ref{sec1} is completely dedicated to a review of the canonical
quantization procedure, in particular in the first paragraph, we explain how
to slice the space-time to arrive to the Arnowitt-Deser-Misner (ADM) form for
the gravitational action, the second paragraph concerns the Wheeler-De Witt
equation (WDE) \cite{DeW1967, DeW1997}, with a brief list of critics moved to
this approach to quantum gravity, which, though consistent, is of course ambiguous.

The main part of our work is, of course, contained in section \ref{sec2},
where we give a detailed explanation of our theory. In paragraph \ref{sec2
nrp} it is treated the very simple case of the quantization of a non
relativistic particle, which introduce to the concept of reparametrization of
the action as a way to extract the right hamiltonian constraint to perform the
quantization. In paragraph \ref{sec2 ka}, in stead, we introduce the
kinematical action \cite{Kuc1981}, giving its physical interpretation
\cite{MerMon2003}. The aim of paragraph \ref{sec2 qf} is to give more insight
into the reparametrization of a classical action in view of the canonical
quantization. In fact, when the scalar field is coupled to the gravitational
one, the dynamics of the background provides automatically the hamiltonian
constraints for the system, but when the background is fixed, in order to
obtain the the right constraints, it is necessary to reparametrize the action
of the scalar field. The reparametrization is performed by the kinematical
action, in the way which has inspired our reformulation of quantum
geometrodynamics described in paragraph \ref{sec2 rq}; postulating the
presence of the kinematical term also in the action we start from in order to quantize the gravitational field \cite{Mon2002}. We show that the hamiltonian operator
is an hermitian one and so the bracket of the quantum states of the system
provides a conserved density of probability during the evolution. The
eigenvalues problem and the smiclassical limit of the theory are faced too.

In section \ref{sec3} we apply our theory to a cosmological
Friedmann-Robertson-Walker model (FRW) obtaining the quantum equation containing also the term due to the density of energy of the dust. After having found a general wave functional for this model, which overcome the not physical initial singularity, replacing it with a more physical peaked density of probability, we give some phenomenological calculations to explore the possibility that the dust be a component of the observed dark matter.

Finally, in the appendix, there is a brief explanation of the so called ``multi time'' approach, which represent another interesting way to arrive to a Schr\"odinger-like quantum dynamics, but profoundly different from the one presented.

\bigskip

\section{Canonical Quantization}

\label{sec1}

The implementation of the canonical quantization formalism to the
gravitational field, leads to the so-called Wheeler-De Witt equation (WDE)
\cite{DeW1967,MisThoWhe1973}, consisting of a functional approach in which the
states of the theory are represented by wave functionals taken on the
3-geometries and, in view of the requirement of general covariance, they do
not possess any real time dependence.\newline Due to its hyperbolic nature,
the WDE is characterized by a large number of unsatisfactory features
\cite{DeW1997}, which strongly support the idea that is impossible any
straightforward extension to the gravitational phenomena of procedures
well-tested only in limited ranges of energies; however in some contexts, like
the very early cosmology \cite{Har1988,KolTur1990} (where a suitable internal
time variable is provided by the volume of the Universe) the WDE is not a
dummy theory and give interesting information about the origin of our
classical universe, see \cite{KirMon1997}, which may be expected to remain
qualitatively valid even for the outcoming of a more consistent approach. In
the following two paragraphs we give a brief review of the canonical method of
quantization for the gravitational field.

\bigskip

\subsection{(3+1)-Slicing Procedure}

\label{sec1 3+1}

To obtain the hamiltonian constraints, which are the starting point for the
canonical quantization of gravity, we have to write the Einstein-Hilbert
action into a (3 + 1) formulation. To this aim, we have to perform a slicing
of the 4-dimensional space-time, on which a metric tensor $g_{\mu\nu}$ is defined.

We consider a space-like hypersurface having a parametric equation $y^{\rho
}=y^{\rho}\left(  x^{i}\right)  $ (Greek labels run from 0 to 3, while Latin
ones run from 1 to 3) and in each point we define a 4-dimensional vector base
composed by its tangent vectors $e_{i}^{\mu}=\partial_{i}y^{\mu}$ and by the
normal unit vector $n^{\mu}$; as just defined, these vectors base satisfy, by
construction, the following relations%
\begin{equation}
g_{\mu\nu}e_{i}^{\mu}n^{\nu}=0,\qquad g_{\mu\nu}n^{\mu}n^{\nu}=-1.
\label{base relation}%
\end{equation}
Now if we deform this hypersurface through the whole space-time, via the
parametric equation $y^{\rho}=y^{\rho}\left(  t,x^{i}\right)  $, we construct
a one-parameter family of space-like hypersurfaces slicing the 4-dimensional
manifold; thus each component of the adapted base, acquiring a dependence on
the time-like parameter $t$, becomes a vector field on the
space-time.\newline Let us introduce the deformation vector $N^{\mu}%
=\partial_{t}y^{\mu}\left(  t,x^{i}\right)  $, which connects two points with
the same spatial coordinates on neighboring hypersurfaces (i.e. corresponding
to values of the parameter $t$ and $t + dt$).\newline This vector field can be
decomposed with respect to the base $\left(  n^{\mu},e_{i}^{\mu}\right)  $,
obtaining the following representation:%
\begin{equation}
N^{\mu}=\partial_{t}y^{\rho}=Nn^{\mu}+N^{i}e_{i}^{\mu}, \label{lap shi dec}%
\end{equation}
where $N$ and $N^{i}$ are, respectively, the lapse function and the shift
vector, so this expression is known as lapse-shift decomposition of the
deformation vector.

It is easy to realize how the space-like hypersurfaces are characterized by
the following 3-dimensional metric tensor $h_{ij}=g_{\mu\nu}e_{i}^{\mu}%
e_{j}^{\nu}$. Since the hypersurface is deformed through space-time, it
changes with a rate, which taken with respect to the label time $t$, can be
decomposed into its normal and tangential contributions%
\begin{equation}
\partial_{t}h_{ij}=-2Nk_{ij}+2\nabla_{(j}N_{i)}, \label{extrinsic curvature}%
\end{equation}
where $N_{i}=h_{ij}N^{j},$ the covariant derivative is constructed with the
3-dimensional metric and $k_{ij}=-\nabla_{i}n_{j}$ denotes the extrinsic curvature.

Now we define the co-base vectors $\left(  n_{\mu},e_{\mu}^{i}\right)  ,$ as
follows%
\begin{equation}
n_{\mu}=g_{\mu\nu}n^{\nu},\qquad e_{\mu}^{i}=h^{ij}g_{\mu\nu}e_{j}^{\nu},
\end{equation}
where $h^{ij}$ is the inverse 3-metric: $h^{ij}h_{jk}=\delta_{k}^{i}.$ By the
second of the above relation we obtain $e_{\mu}^{i}e_{j}^{\mu}=\delta_{j}^{i}.$

The explicit expression for the 4-metric tensor $g_{\mu\nu}$ and its inverse
$g^{\mu\nu}$ assume, in the system $\left(  t,x^{i}\right)  $, respectively,
the form:%
\begin{equation}
g_{\mu\nu}=\left(
\begin{array}
[c]{cc}%
N_{i}N^{i}-N^{2} & N_{i}\\
N_{i} & h_{ij}%
\end{array}
\right)  ,\qquad g^{\mu\nu}=\left(
\begin{array}
[c]{cc}%
-\dfrac{1}{N^{2}} & \dfrac{N^{i}}{N^{2}}\\
\dfrac{N^{i}}{N^{2}} & h^{ij}-\dfrac{N^{i}N^{j}}{N^{2}}%
\end{array}
\right)  .\label{metric}%
\end{equation}
Moreover, the normal unit vector $n^{\mu}$ has the following components
$\left(  \dfrac{1}{N},-\dfrac{N^{i}}{N}\right)  $, and this implies that the
covariant normal vector be $n_{\mu}=\left(  -N,0\right)  $; below we will use
to indicate the components of the vectors in the system $\left(
t,x^{i}\right)  ,$ with Greek barred labels as: $\bar{\mu},$ $\bar{\nu}$,
$\bar{\rho}.....$ We also note that in this system of coordinates, the square
root of the determinant of the metric tensor assumes the form $\sqrt
{-g}=N\sqrt{h}.$

It is possible to show that the Einstein-Hilbert action can be rewritten as
follows \cite{ArnDesMis1962, Wal1984, Thi2001}:%
\begin{equation}
S=\int\limits_{\Sigma^{3}\times\Re}dtd^{3}xN\sqrt{h}\left(  ^{\left(
3\right)  }R+k_{ij}k^{ij}-k^{2}\right)  ,\label{azione grav}%
\end{equation}
which is the most appropriate to construct the ``ADM action'' for the
gravitational field.\newline Now, defining the conjugate momenta to the
dynamical variables, which are the component of the 3-metric tensor, we can
rewrite the gravitational action in its hamiltonian form. The gravitational
Lagrangian $L^{g}$ does not contain the time derivative of the lapse function
$N$ and of the shift vector $N^{i},$ so their conjugate momenta are
identically zero and the Lagrangian is said singular. Summarizing, we have for
the conjugate momenta:%
\begin{align}
&  p^{ij}\left(  t,x^{i}\right)  =\dfrac{\partial L^{g}}{\partial\left(
\partial_{t}h_{ij}\right)  }=\sqrt{h}\left(  k^{ij}-kh^{ij}\right)  ,\\
\pi\left(  t,x^{i}\right)   &  =\dfrac{\partial L^{g}}{\partial\left(
\partial_{t}N\right)  }=0,\qquad\pi_{i}\left(  t,x^{i}\right)  =\dfrac
{\partial L^{g}}{\partial\left(  \partial_{t}N^{i}\right)  }=0.
\end{align}
\qquad By the above definition, we can perform the Legendre dual
transformation and, with few algebra, then obtaining the below final form for
the gravitational action \cite{Thi2001}:%
\begin{align}
S^{g}\left(  h_{ij},p^{kl},N,N^{a},\pi,\pi_{b}\right)   &  =\int
\limits_{\Sigma^{3}\times\Re}dtd^{3}x\left\{  p^{ij}\partial_{t}h_{ij}%
+\pi\partial_{t}N+\pi_{k}\partial_{t}N^{k}\right.  \nonumber\\
&  -\left.  \left(  \lambda\pi+\lambda^{j}\pi_{j}+NH^{g}+N^{i}H_{i}%
^{g}\right)  \right\}  ,\label{az grav}%
\end{align}
where the so called super-hamiltonian $H^{g}$ and super-momentum $H_{i}^{g},$
read respectively as%
\begin{equation}
H^{g}=G_{ijkl}p^{ij}p^{kl}-\sqrt{h}^{\left(  3\right)  }R,\qquad H_{i}%
^{g}=-2\nabla_{j}p_{i}^{j},\label{manu}%
\end{equation}
where (using geometrical units) $G_{ijkl}=\dfrac{1}{2\sqrt{h}}\left(
h_{ik}h_{jl}+h_{il}h_{jk}-h_{ij}h_{kl}\right)  $ is the so-called super-metric
(Wheeler 1968).

Now, before calculating the other dynamical equations, we want to add to this
picture, also a matter field, which, for simplicity, is represented by a
self-interacting scalar field $\phi.$ This lead us to the following expression
for the action of the gravitational and matter field:%
\begin{align}
S^{g\phi} &  =\int\limits_{\Sigma^{3}\times\Re}dtd^{3}x\left\{  p^{ij}%
\partial_{t}h_{ij}+\pi\partial_{t}N+\pi_{k}\partial_{t}N^{k}+p_{\phi}%
\partial_{t}\phi\right.  \nonumber\\
&  -\left.  \left(  \lambda\pi+\lambda^{j}\pi_{j}+N\left(  H^{g}+H^{\phi
}\right)  +N^{i}\left(  H_{i}^{g}+H_{i}^{\phi}\right)  \right)  \right\}
\label{grav-scal action}%
\end{align}
where the hamiltonian terms $H^{\phi}$ and $H_{i}^{\phi}$ read explicitly as:%
\begin{equation}
H^{\phi}=\dfrac{1}{2\sqrt{h}}p_{\phi}^{2}+\dfrac{\sqrt{h}}{2}h^{ij}%
\partial_{i}\phi\partial_{j}\phi+\sqrt{h}V\left(  \phi\right)  \qquad
H_{i}^{\phi}=p_{\phi}\partial_{i}\phi\label{manu1}%
\end{equation}
where $h=\det\{h_{ij}\}$ and $V\left(  \phi\right)  $ denotes a
self-interaction potential energy.

Varying the action (\ref{grav-scal action}) with respect to the Lagrange
multipliers $\lambda$ and $\lambda_{i}$, we obtain the first class
constraints:%
\begin{equation}
\pi=0,\qquad\pi_{k}=0;\label{ham con}%
\end{equation}
to assure that the dynamics be consistent, the Poisson parentheses, between the
constraints and the hamiltonian, have to be zero, so we must require that the
second class constraints%
\begin{equation}
H^{g}+H^{\phi}=0,\qquad H_{i}^{g}+H_{i}^{\phi}=0,\label{ham con2}%
\end{equation}
be satisfied.\newline Moreover varying the action with respect the two
conjugate momenta $\pi$ and $\pi_{i}$, we obtain the two equations:%
\begin{equation}
\partial_{t}N=\lambda,\qquad\partial_{t}N^{i}=\lambda^{i},\label{eq lap shi}%
\end{equation}
which assure that the trajectories of the lapse function and of the shift
vector in the phase space are completely arbitrarily.

The action (\ref{grav-scal action}) has to be varied with respect to all the
dynamical variables and this gives us the hamiltonian equations for the scalar
and gravitational field, which take the following form:%
\begin{equation}
\dfrac{d}{dt}h_{ab}=2NG_{abkl}p^{kl}+2\nabla_{(a}N_{b)}, \label{eq per h}%
\end{equation}%
\begin{align}
\dfrac{d}{dt}p^{ab}  &  =\dfrac{1}{2}N\dfrac{h^{ab}}{\sqrt{h}}\left(
p^{ij}p_{ij}-\dfrac{1}{2}p^{2}\right)  -\dfrac{2N}{\sqrt{h}}\left(
p^{ai}p_{i}^{b}-\dfrac{1}{2}pp^{ab}\right)  +\nonumber\\
&  -N\sqrt{h}\left(  ^{\left(  3\right)  }R^{ab}-\dfrac{1}{2}^{\left(
3\right)  }Rh^{ab}\right)  +\nonumber\\
&  +\sqrt{h}\left(  \nabla^{a}\nabla^{b}N-h^{ab}\nabla^{i}\nabla_{i}N\right)
+\nonumber\\
&  -2\nabla_{i}\left(  p^{i(a}N^{b)}\right)  +\nabla_{i}\left(  N^{i}%
p^{ab}\right)  +\nonumber\\
&  +\dfrac{N}{4\sqrt{h}}h^{ab}p_{\phi}^{2}-\dfrac{N}{2}\sqrt{h}h^{ab}\left(
\dfrac{1}{2}h^{ij}\partial_{i}\phi\partial_{j}\phi+V\left(  \phi\right)
\right)  , \label{eq per p}%
\end{align}%
\begin{equation}
\dfrac{d}{dt}\phi=\dfrac{N}{\sqrt{h}}p_{\phi}+N^{i}\partial_{i}\phi,
\end{equation}%
\begin{align}
\dfrac{d}{dt}p_{\phi}  &  =N\sqrt{h}h^{ij}\partial_{i}\partial_{j}%
\phi+\partial_{j}\left(  N\sqrt{h}h^{ij}\right)  \partial_{i}\phi+\nonumber\\
&  -N\sqrt{h}\dfrac{\partial V\left(  \phi\right)  }{\partial\phi}%
+\partial_{i}\left(  N^{i}p_{\phi}\right)  .
\end{align}
The complete dynamics of the coupled gravito-scalar system is represented by
the above dynamical equations together with equation (\ref{eq lap shi}) and
the first and second class constraints (\ref{ham con}) and (\ref{ham con2}),
which tell us we can not choose the fields and their conjugate momenta arbitrarily.

\bigskip

\subsection{ADM Action and Wheeler-De Witt Equation}

\label{sec1 WDE}

Now we briefly recall how the Wheeler-De Witt approach \cite{DeW1967, Kuc1981}
faces the problem of quantizing a coupled system consisting of gravity and a
real scalar field, which implies also the metric field now be a dynamical
variable. The action describing this coupled system reads%
\begin{align}
S^{g\phi} &  =\int\limits_{\Sigma^{3}\times\Re}dtd^{3}x\left\{  p^{ij}%
\partial_{t}h_{ij}+\pi\partial_{t}N+\pi_{k}\partial_{t}N^{k}+p_{\phi}%
\partial_{t}\phi\right.  \nonumber\\
&  -\left.  \left(  \lambda\pi+\lambda^{j}\pi_{j}+N\left(  H^{g}+H^{\phi
}\right)  +N^{i}\left(  H_{i}^{g}+H_{i}^{\phi}\right)  \right)  \right\}
,\label{s}%
\end{align}
where $p^{ij}$ denotes the conjugate momenta to the 3-dimensional metric
tensor $h_{ij}$ and the super-hamiltonian and super-momentum terms take the
form contained in equations (\ref{manu}) e (\ref{manu1}). 

Since now $N$ and $N^{i}$ are, in principle, dynamical variables, they have to
be varied, so leading to the constraints $H^{g}+H^{\phi}=0$ and $H_{i}%
^{g}+H_{i}^{\phi}=0$ which are equivalent to the $\mu-0$-components of the
Einstein equations and therefore play the role of constraints for the Cauchy
data. It is just this restriction on the initial values problem, a peculiar
difference between the previous case, at fixed background, and the present
one; in fact, now, on the regular hypersurface $t=t_{0}$, the initial
conditions $\{\phi_{0}(x^{i}),\;p_{0}(x^{i}),\;h_{ij\;0}(x^{i}),\;p^{kl}%
(x^{i})\}$ can not be assigned freely, but they must verify on $\Sigma_{t_{0}%
}^{3}$ the four relations $\left(  H^{g}+H^{\phi}\right)  \mid_{t_{0}}=0$ and
$\left(  H_{i}^{g}+H_{i}^{\phi}\right)  \mid_{t_{0}}=0$.

Indeed behaving like Lagrange multipliers, the lapse function and the shift
vector have not a real dynamics and their specification corresponds to assign
a particular slicing of $\mathcal{M}^{4}$, i. e. a system of reference.

In order to quantize this system we assume that its states be represented by a
wave functional $\Psi(N,N^{k},h_{ij},\phi)$ and implement the canonical
variables to operators acting on this wave functional (in particular we set
$h_{ij}\rightarrow\widehat{h}_{ij},\;p^{ij}\rightarrow\widehat{p}^{ij}%
\equiv-i\hbar\delta(\;)/\delta h_{ij}$).\newline The quantum dynamics of the
system is then induced by imposing the operators translation of the
classical constraints, which leads to the following quantum equations:%
\begin{align}
\widehat{\pi}\Psi & =0,\qquad\widehat{\pi}_{k}\Psi=0,\nonumber\\
(\widehat{H}_{i}^{g}+\widehat{H}_{i}^{\phi})\Psi & =0,\,\quad(\widehat{H}%
^{g}+\widehat{H}^{\phi})\Psi=0,\label{v}%
\end{align}
which to be solved it would require a specific choice for the \emph{normal
ordering} of the operators. The first seven quantum equations can be simply
solved: they restrict the dependence of the wave functional only on a class of
3-geometries, which we indicate with $\left\{  h_{ij}\right\}  .$ The last one
is the Wheeler-De Witt equation, which, in view of what just said, we rewrite
$(\widehat{H}^{g}+\widehat{H}^{\phi})\Psi\left(  \phi,\left\{  h_{ij}\right\}
\right)=0.$

Due to its hyperbolic nature this formulation of the quantum dynamics has some
limiting feature \cite{DeW1997} , which we summarize by the following three
points:\newline i) It does not exist any general procedure allowing to turn the
space of the solutions into an Hilbert one and so any appropriate general
notion of functional probability distribution is prevented.\newline ii) The
WDE does not contain any dependence on the variable $t$ or on the function
$y^{\mu}$, so loosing its evolutive character along the slicing $\Sigma
_{t}^{3}$. Moreover individualizing an internal variable which can play the
role of ``time '' is an ambiguous procedure which does not lead to a general
prescription.\newline iii)At last we stress what is to be regarded as an
intrinsic inconsistency of the approach above presented: the WDE is based on
the primitive notion of space-like hypersurfaces, i.e. of a time-like normal
field, which is in clear contradiction with the random behavior of a quantum
metric field \cite{ZinJus1996}; indeed the space or time character of a vector
becomes a precise notions only in the limit of a perturbative quantum gravity
theory. This remarkable ambiguity leads us to infer that there is
inconsistency between the requirement of a wave equation (i.e. a wave
functional) invariant, like the WDE one, under space diffeomorphisms and time
displacements on one hand, and, on the other one, the (3 + 1)-slicing
representation of the global manifold.

The existence of these shortcomings in the WDE approach, induces us to search
for a better reformulation of the quantization procedure which addresses the
solution of the above indicated three points as prescriptions to write down
new dynamical quantum constraints.

\bigskip

\section{Reformulation of Quantum Dynamics}

\label{sec2}

Our reformulation of the canonical quantum gravity is based on a fundamental
criticism about the possibility to speak of a unit time-like normal field and
of space-like hypersurfaces, which are at the ground of the ADM formalism,
when referring to a quantum space-time; in fact, in this case, either the
time-like nature of a vector field, either the space-like nature of the
hypersurfaces can be recognized at most in average sense, i.e. with respect to
expectation values. This consideration makes extremely ambiguous to apply the
(3+1)-splitting on a quantum level and leads us to claim that the canonical
quantization of gravity has sense only when referred to a fixed slicing, or in
other words, when referred to a fixed reference frame, i.e. only after the
notion of space and time are physically distinguishable. To fix the slicing we
have to choose a particular family of hypersurfaces and this means we have to
fix the lapse function $N$ and the shift vector $N^{i}.$ However, so doing, we
loose the hamiltonian constraints (\ref{ham con}), (\ref{ham con2}) and, with
them, the standard procedure to quantize the dynamics of the system; as a
solution to this problem, we propose to reparametrize the gravitational action
using the so called Kinematical Action, obtaining new hamiltonian constraints
and going on toward the canonical quantization of the system.

\bigskip

\subsection{Non Relativistic Particle}

\label{sec2 nrp}

As an helpful example for the analysis below developed, we review the case of
the one-dimensional non-relativistic (parametrized) particle, whose action reads%
\begin{equation}
S=\int\{p\dot{q}-h(p,q)\}dt\,,\label{a}%
\end{equation}
where $t$ denotes the Newton time and $h$ the hamiltonian function. In order
to quantize this system, we parameterize the Newton time as $t=t(\tau)$, so
getting the new action as%
\begin{equation}
S=\int\{p\frac{dq}{d\tau}-h(p,q)\frac{dt}{d\tau}\}d\tau\,.\label{b}%
\end{equation}
Now we set $p_{0}\equiv-h$ and add this relation to the above action by a
Lagrangian multiplier $\lambda$, i.e.%
\begin{equation}
S=\int\{p\frac{dq}{d\tau}+p_{0}\frac{dt}{d\tau}-h(p,q,p_{0},\lambda
)\}d\tau\,\quad h\equiv\lambda(h+p_{0})\,.\label{c2}%
\end{equation}
By varying this action with respect to $p$ and $q$, we get the Hamilton
equations $dq/d\tau=\lambda\partial h/\partial p$ and $dp/d\tau=-\lambda
\partial h/\partial q$, while the variations of $p_{0}$ and $t$ yield
$dt/d\tau=\lambda$ and $dp_{0}/d\tau=0$; all together, these equations
describe the same Newton dynamics, having the energy as constant of the
motion. But now, by varying $\lambda$, we get the (desired) constraint
$h+p_{0}=0$, which, in terms of the operators $\widehat{p}_{0}=-i\hbar
\partial_{t}$ and $\widehat{h}$, provides the Schr\"{o}dinger equation
$i\hbar\partial_{t}\psi=\widehat{h}\psi$, as taken for the system state
function $\psi(t,q)$. Finally we remark that, when retaining the relation
$dt/d\tau=\lambda$, we are able to write the wave equation in the parametric
time as%
\begin{equation}
i\hbar\partial_{\tau}\psi(\tau,q)=\lambda(\tau)\widehat{h}\psi(\tau
,q)\,,\label{d1}%
\end{equation}
where $\lambda(\tau)$ is to be assigned. 

In spite of its simplicity, this example is a naive, but very good prototype
of our approach to the canonical quantum gravity.

\bigskip

\subsection{Kinematical Action and its Physical Interpretation}

\label{sec2 ka}

We have introduced in the previous section the lapse-shift decomposition of
the deformation vector (\ref{lap shi dec}). It is worth noting that we can
obtain such equation varying an action built to this aim. It is the so-called
kinematical action and takes the following form:%
\begin{equation}
S=\int\limits_{\Sigma^{3}\times\Re}dtd^{3}x\left(  p_{\mu}\partial_{t}y^{\mu
}-Np_{\mu}n^{\mu}-N^{i}p_{\mu}e_{i}^{\mu}\right)  .\label{kin act}%
\end{equation}
If we now vary the action (\ref{kin act}) with respect to the dynamical
variables $p_{\mu}$ and $y^{\mu}$, and we put these two variations equal to
zero, we obtain respectively:%
\begin{equation}
\partial_{t}y^{\mu}=Nn^{\mu}+N^{i}\partial_{i}y^{\mu},\qquad\partial_{t}%
p_{\mu}=-Np_{\rho}\partial_{\mu}n^{\rho}+\partial_{i}\left(  N^{i}p_{\mu
}\right)  .\label{eq din}%
\end{equation}
The first one of such equations is the lapse-shift decomposition of the
deformation vector, while the second one provides the dynamical evolution for
$p_{\mu},$ which is the conjugate momenta to the vector $y^{\rho}.$

The kinematical action is used in quantum field theory on curved space-time,
in order to reparameterize the field action \cite{BirDav1982, Kuc1981}, but it
will be clear in the next section how, in our approach, it plays an important
role also in the reformulation of the canonical quantum gravity.

In this section we want to investigate the physical meaning of the
``kinematical term'', which will outline either the main aspects of our
reformulation of the canonical quantum gravity, either the meaning of the
reparameterization in quantum field on curved space.

To get the searched physical insight, let us rewrite the equations (\ref{eq
din}) in a covariant form. To this aim we recall to denote the coordinates
$(t,x^{i})$ by barred Greek labels: $\overline{\mu},\overline{\nu}%
,\overline{\rho}....$ and we also remark that the following relations take
place: $\partial_{t}=\partial_{t}y^{\mu}\partial_{\mu},$ $\partial
_{i}=\partial_{i}y^{\mu}\partial_{\mu},$\ $n^{\overline{\mu}}\partial
_{\overline{\mu}}=n^{\mu}\partial_{\mu}.$\newline Now remembering that the
normal vector $n^{\mu}$ has components $n^{\overline{\mu}}\equiv\left(
\dfrac{1}{N},-\dfrac{N^{i}}{N}\right)  $ in the system $\left(  t,x^{i}%
\right)  ,$ it is possible to rewrite the first one of equations (\ref{eq
din}) in the following form: $n^{\mu}=n^{\overline{\rho}}\partial
_{\overline{\rho}}y^{\mu};$ this equation ensures that, after the variation
$n^{\mu}$ is a real unit time-like vector, i.e.%
\begin{equation}
g_{\mu\nu}n^{\mu}n^{\nu}=g_{\mu\nu}n^{\overline{\rho}}\partial_{\overline
{\rho}}y^{\mu}n^{\overline{\sigma}}\partial_{\overline{\sigma}}y^{\nu
}=g_{\overline{\rho}\overline{\sigma}}n^{\overline{\rho}}n^{\overline{\sigma}%
}=-1,\label{eq nor vec}%
\end{equation}
the last equality being true by construction of $g_{\overline{\mu}%
\overline{\nu}}$ and $n^{\overline{\mu}}.$ Moreover, since $n^{\mu}$ is in any
system of coordinates normal to the hypersurfaces $\Sigma^{3},$ then we see
how the use of the kinematical action allows to overcome the ambiguity in the
existence of a real time-like normal vector field, we have spoken about in the
introduction of this paper.\newline Now using the relations $\partial
_{t}=\partial_{t}y^{\mu}\partial_{\mu},$ $\partial_{i}=\partial_{i}y^{\mu
}\partial_{\mu},$\ $n^{\overline{\mu}}\partial_{\overline{\mu}}=n^{\mu
}\partial_{\mu}$ and the first one of equations (\ref{eq din}), we may rewrite
the second kinematical equation, concerning the momentum dynamics as follows:%
\begin{equation}
n^{\rho}\left[  \partial_{\rho}\left(  Np_{\mu}\right)  -\partial_{\mu}\left(
Np_{\rho}\right)  \right]  =-\partial_{\mu}\left(  Np_{\nu}n^{\nu}\right)
+p_{\mu}\left(  n^{\rho}\partial_{\rho}N+\partial_{i}N^{i}\right)
;\label{eq4}%
\end{equation}
we note that $p_{\mu}$ is not a vector, but it is a vector density of weight
1/2; thus we can rewrite it as $p_{\mu}=-\sqrt{h}\varepsilon\pi_{\mu}$, where
$\varepsilon$ is a real 3-scalar and $\pi_{\mu}$ is a vector, such that it
satisfies the relation $n^{\mu}\pi_{\mu}=-1.$ Using this new expression for
$p_{\mu}$, equation (\ref{eq4}) rewrites:%
\begin{equation}
\varepsilon n^{\rho}\left(  \partial_{\rho}\pi_{\mu}-\partial_{\mu}\pi_{\rho
}\right)  =-\pi_{\mu}\frac{1}{\sqrt{-\overline{g}}}\partial_{\rho}\left(
\sqrt{-\overline{g}}\varepsilon n^{\rho}\right)  ,\label{eq5}%
\end{equation}
which covariantly reads%
\begin{equation}
\varepsilon n^{\rho}\left(  \nabla_{\rho}\pi_{\mu}-\nabla_{\mu}\pi_{\rho
}\right)  +\pi_{\mu}\nabla_{\rho}\left(  \varepsilon n^{\rho}\right)
=0.\label{eq6}%
\end{equation}
Then, multiplying equation (\ref{eq6}) for $n^{\mu}$, we get%
\begin{equation}
\nabla_{\rho}\left(  \varepsilon n^{\rho}\right)  =0.\label{eq con}%
\end{equation}
A perfect fluid, with entropy density $\sigma$ and velocity $u_{\mu},$
satisfies the equation $\nabla_{\mu}\left(  \sigma u^{\mu}\right)  =0$, but
for a dust case the density of entropy is proportional to the density of
energy $\left(  \sigma\propto\varepsilon\right)  $, so that equation (\ref{eq
con}) is the one for a dust fluid of density of energy $\varepsilon$ and
4-velocity $n_{\mu}.$\newline Now, using equation (\ref{eq con}), we can
rewrite the relation (\ref{eq6}) as%
\begin{equation}
n^{\rho}\left(  \nabla_{\rho}\pi_{\mu}-\nabla_{\mu}\pi_{\rho}\right)
=0.\label{eq7}%
\end{equation}
Setting now $\pi_{\mu}=n_{\mu}+s_{\mu}$, with $n^{\mu}s_{\mu}=0$, from above,
we arrive to%
\begin{equation}
n^{\rho}\nabla_{\rho}n_{\mu}=n^{\rho}\left(  \nabla_{\mu}s_{\rho}-\nabla
_{\rho}s_{\mu}\right)  =\gamma n^{\rho}F_{\mu\rho},\label{eq8}%
\end{equation}
with $s_{\rho}=\gamma A_{\rho}$, where $\gamma$ is a constant and $F_{\mu\rho
}=\nabla_{\mu}A_{\rho}-\nabla_{\rho}A_{\mu}$ (obviously $n^{\rho}A_{\rho}%
=0$).\newline Thus equation (\ref{eq8}), together with (\ref{eq con}) are the
field equations of a dust fluid with density of energy $\varepsilon$, whose
4-velocity $n^{\mu}$ is tangent to a space-time curve associated to the
presence of an ``electromagnetic-like'' field (say a
\emph{gravito-electromagnetic field}). So, on a classical level, the
kinematical action is equivalent to the action of such a dust fluid and, in
this sense, it is upgraded from its geometrical nature to a physical one.

The condition $n^{\rho}A_{\rho}=0$ can be written in the system $\left(
t,x^{i}\right)  $ as $n_{\overline{\rho}}A^{\overline{\rho}}=0$, from which it
follows $A^{\overline{0}}=0$ and this means that in the fluid reference we
have to do with a gauge condition such that $A^{\mu}\equiv\left(
0,\underline{A}\right)  ,$ i.e. with a simple 3-vector potential for the
gravito-electromagnetic field.

Now let us come back to the kinematical action (\ref{kin act}): the
corresponding super-hamiltonian and super-momentum of the kinematical term are:%
\begin{equation}
H^{k}=p_{\mu}n^{\mu},\qquad H_{i}^{k}=p_{\mu}e_{i}^{\mu},\label{sup kin}%
\end{equation}
Using the definitions above introduced for $p_{\mu}$ and $s_{\mu}$ we have:%
\begin{equation}
H^{k}=\sqrt{h}\varepsilon,\qquad H_{i}^{k}=-\sqrt{h}\varepsilon\gamma A_{\mu
}e_{i}^{\mu}.\label{ham mom}%
\end{equation}
It is clear that $A_{\mu}e_{i}^{\mu}=A_{\mu}\dfrac{\partial y^{\mu}}{\partial
x^{i}}$ is a transformation of coordinates from the generic system $y^{\mu}$
to the system of the hypersurface, that is the one which we have before
indicated with barred labels. So we write $A_{\mu}e_{i}^{\mu}=A_{i}$, that is
we introduce the projection of the field $A_{\mu}$ on the spatial
hypersurfaces.\newline So equations (\ref{ham mom}) rewrites as:%
\begin{equation}
H^{k}=\sqrt{h}\varepsilon,\qquad H_{i}^{k}=-\sqrt{h}\varepsilon\gamma A_{i}.
\label{ham mom2}%
\end{equation}
In \cite{Mon2002} is shown that the energy-momentum tensor of the dust is
orthogonal to the hypersurfaces $\Sigma^{3};$ this is the reason why such tensor contributes only to the super-hamiltonian constraint, 
by its energy density term. Moreover, it
is possible to show, via a simple model, why the presence of the field
$A_{\mu}$ has, instead, effects only on the super-momentum. To this end, let
us consider an interaction between a current $j^{\mu}$ and a field $B_{\mu}$;
then the hamiltonian of interaction will be:%
\begin{equation}
H_{int}=\int d^{4}x\sqrt{-g}j^{\mu}B_{\mu}.\label{ham int}%
\end{equation}
Since $H_{int}$ is obviously a scalar, we can rewrite it in the system of
coordinates with barred labels, as follows%
\begin{equation}
H_{int}=\int d^{4}\overline{x}N\sqrt{h}j^{\overline{\mu}}B_{\overline{\mu}%
};\label{ham int bar}%
\end{equation}
taking now $j^{\overline{\mu}}=\varepsilon n^{\overline{\mu}}$ (current of
matter) and $B_{\overline{\mu}}=\gamma A_{\overline{\mu}},$ we have,
remembering also that $n^{\overline{\mu}}\equiv\left(  \dfrac{1}{N}%
,-\dfrac{N^{i}}{N}\right),$%
\begin{equation}
H_{int}=\int d^{4}\overline{x}\sqrt{h}\varepsilon\gamma\left(  A_{\overline
{0}}-N^{\overline{i}}A_{\overline{i}}\right)  .\label{ham int fin}%
\end{equation}
This expression no more depends on the lapse function, so that it does not
contribute to the super-hamiltonian, while the contribution to the
super-momentum is just the one in equation (\ref{ham mom}).

Above we have introduced the projection of the field $A_{\mu}$ on the spatial
hypersurfaces, i.e. $A_{i}=A_{\mu}e_{i}^{\mu};$ this is of course a simple
transformation of coordinates, but it does not assure $A_{i}$ is a 3-vector.
To show this, we define $A^{i}=A^{\mu}e_{\mu}^{i};$ it is worth noting that it
is not a transformation of coordinates, but this choice on how to project the
contravariant 4-vector $A^{\mu}$, is sufficient to show that $A_{i}%
=h_{ik}A^{k},$ which ensures $A_{i}$ is a 3-vector on the hypersurfaces, which
lowers and raises its index by the induced 3-metric.\newline In fact starting
from the expression of $A_{i}$ and recalling that $e_{i}^{\mu}=h_{ik}g^{\mu
\nu}e_{\nu}^{k},$, we can write:%
\begin{equation}
A_{i}=A_{\mu}e_{i}^{\mu}=A_{\mu}h_{ik}g^{\mu\nu}e_{\nu}^{k}=h_{ik}A^{\nu
}e_{\nu}^{i}=h_{ik}A^{k},\label{3-vector}%
\end{equation}
where, in the last equality, we have used the definition of $A^{k}.$

To conclude this section, we want to study the behaviors of $\varepsilon$ and
$A_{i};$ to this end we start from equations (\ref{eq din}), multiplying the
second one by $n^{\mu},$ and remembering that $n^{\mu}\partial_{\mu
}=n^{\overline{\mu}}\partial_{\overline{\mu}},$ we arrive to%
\begin{equation}
\partial_{t}\left(  \sqrt{h}\varepsilon\right)  -\partial_{i}\left(  \sqrt
{h}\varepsilon N^{i}\right)  =0;\label{eq e}%
\end{equation}
Moreover, by multiplying the second one with $e_{i}^{\mu}$ and considering
also the first kinematical equation, we get an expression of the form:%
\begin{equation}
\partial_{t}\left(  \sqrt{h}\varepsilon\gamma A_{i}\right)  -\partial
_{k}\left(  \sqrt{h}\varepsilon\gamma N^{k}A_{i}\right)  =\sqrt{h}%
\varepsilon\gamma A_{k}\partial_{i}N^{k}-\sqrt{h}\varepsilon\partial
_{i}N.\label{eq A}%
\end{equation}
To treat these two equations (\ref{eq e}) and (\ref{eq A}) in a general
reference frame, it is a very difficult task, but it becomes very simple in a
synchronous reference, where $N=1$ and $N^{i}=0;$ in this particular case we have:%
\begin{equation}
\partial_{t}\left(  \sqrt{h}\varepsilon\right)  =0,\qquad\partial_{t}\left(
\sqrt{h}\varepsilon\gamma A_{i}\right)  =0.
\end{equation}
The first one of the above equations means that $\sqrt{h}\varepsilon
=-\omega\left(  x^{i}\right)  $ where $\omega$ is a scalar density of weight
1/2, which depends only on $x^{i};$ we note that $\varepsilon=-\dfrac
{\omega\left(  x^{i}\right)  }{\sqrt{h}}$, this means $\varepsilon$ is the
density of energy of a non relativistic dust. While from the second one we
obtain $\gamma A_{i}\omega\left(  x^{i}\right)  =-k_{i}\left(  x^{k}\right)
,$ which is a 3-vector density of weight 1/2 and depends only on $x^{i}$ (we
have to do with a simple magnetic term). It is clear that we can now write the
super-hamiltonian and super-momentum of the kinematical term as follows%
\begin{equation}
H^{k}=-\omega\left(  x^{l}\right)  ,\qquad H_{i}^{k}=-k_{i}\left(
x^{l}\right)  .\label{raffaella}%
\end{equation}
We will return on the above expression in the next section, when treating the
eigenvalues problem and the classical limit of the quantized theory; indeed we
will find a connection between the density of energy of the dust and the
eigenvalue of the super-hamiltonian operator as well as between the
eigenvalues of the super-momentum operator and the presence of the field
$A_{i}.$

\bigskip

\subsection{Quantum Fields on Curved Background}

\label{sec2 qf}

Now, within the ADM formalism, we analyze the quantization of a
self-interacting scalar field $\phi(t, x^{i})$ described by a potential term
$V(\phi)$ on a fixed gravitational background; its dynamics is summarized by
the action%
\begin{equation}
S^{\phi}(\pi_{\phi},\phi)=\int_{\mathcal{M}^{4}}\left\{  p_{\phi}\partial
_{t}\phi-NH^{\phi}-N^{i}H_{i}^{\phi}\right\}  d^{3}xdt,\label{c}%
\end{equation}
where $p_{\phi}$ denotes the conjugate field to the scalar one and the
hamiltonian terms $H^{\phi}$ and $H_{i}^{\phi}$ are those contained in
equations (\ref{manu1}).\newline This action should be varied with respect to
$p_{\phi}$ and $\phi$, but not $N$, $N^{i}$ and $h^{ij}$ since the metric
background, in this case, is assigned; but if we want to apply to this system
the canonical quantization formalism we have to extract the hamiltonian
constraints by a reparametrization of the action for the scalar field. This
aim is reached by adding to $S^{\phi}$ the kinematical action (\ref{kin act}),
moreover, this additional term has a geometrical as well as a physical
interpretation as seen above.The total action is%
\begin{equation}
S^{\phi k}\equiv S^{\phi}+S^{k}=%
%TCIMACRO{\dint \limits_{\mathcal{M}^{4}}}%
%BeginExpansion
{\displaystyle\int\limits_{\mathcal{M}^{4}}}
%EndExpansion
\left\{  p_{\phi}\partial_{t}\phi+p_{\mu}\partial_{t}y^{\mu}-N(H^{\phi}%
+H^{k})-N^{i}(H_{i}^{\phi}+H_{i}^{k})\right\}  d^{3}xdt,\label{f}%
\end{equation}
In the above action $n^{\mu}$ and $h^{ij}$ are to be regarded as assigned
functionals of $y^{\mu}(t,x^{i})$; background is now fixed by the
hypersurfaces $y^{\mu}$ and their normal vector $n^{\mu}$, so we can consider
$N$ and $N^{i}$ as generic Lagrange multipliers, the addition of the kinematical action does
not affect the field equation for the scalar field, while the variations with
respect to $p_{\mu}$ and $y^{\mu}$ provide the equation (\ref{eq din}) and the
evolution of the kinematical momentum.\newline Finally, by varying, now even,
with respect to $N$ and $N^{i}$ we get the  constraints%
\begin{equation}
H^{\phi}=-p_{\mu}n^{\mu},\,\quad H_{i}^{\phi}=-p_{\mu}e_{i}^{\mu},\label{i}%
\end{equation}
Clearly is to be assigned the following Cauchy problem assigned on a regular
initial hypersurface $\Sigma_{t_{0}}^{3}$, i. e. $y^{\mu}(t_{0},x^{i}%
)=y_{0}^{\mu}(x^{i})$%
\begin{align}
\phi(t_{0},x^{i})  & =\phi_{0}(x^{i}),\,\quad\pi_{\phi}(t_{0},x^{i})=\pi
_{0}(x^{i}),\,\quad\nonumber\\
y^{\phi}(t_{0},x^{i})  & =y_{0}^{\mu}(x^{i}),\,\quad p_{\mu}(t_{0}%
,x^{i})=p_{\mu\;0}(x^{i}),\label{l}%
\end{align}
At last, to complete the scheme of the field equations, we have also to
specify the lapse function and the shift vector by the first of equations
(\ref{eq din}), but also the metric tensor $h_{ij}$ by the relation
$h_{ij}=g_{\mu\nu}\partial_{i}y^{\mu}\partial_{j}y^{\nu}$.

This system can be easily quantized in the canonical formalism by assuming the
states of the system be represented by a wave functional $\Psi(y^{\mu}%
(x^{i}),\phi(x^{i}))$ and implementing the canonical variables $\{y^{\mu
},p_{\mu},\phi,p_{\phi}\}$ to operators $\{\widehat{y}^{\mu},\;\widehat
{p}_{\mu}=-i\hbar\delta(\;)/\delta y^{\mu},\;\widehat{\phi},\;\widehat
{p}_{\phi}=-i\hbar\delta(\;)/\delta\phi\}$. Then the quantum dynamics is
described by the equations%
\begin{align}
i\hbar n^{\mu}\frac{\delta\Psi}{\delta y^{\mu}}=\widehat{H}^{\phi}\Psi &
=\left[  -\frac{\hbar^{2}}{2\sqrt{h}}\frac{\delta}{\delta\phi}\frac{\delta
}{\delta\phi}+\frac{1}{2}\sqrt{h}h^{ij}\partial_{i}\phi\partial_{j}\phi
+\sqrt{h}V(\phi)\right]  \Psi\,,\quad\label{m}\\
i\hbar e_{i}^{\mu}\frac{\delta\Psi}{\delta y^{\mu}} &  =\widehat{H}_{i}^{\phi
}\Psi=-i\hbar\partial_{i}\phi\frac{\delta\Psi}{\delta\phi}.\label{m1}%
\end{align}
These equations have $5\times\infty^{3}$ degrees of freedom, corresponding to
the values taken by the four components of $y^{\mu}$ and the scalar field
$\phi$ in each point of a spatial hypersurface. In (\ref{m}) and (\ref{m1})
$y^{\mu}$ plays the role of ``time variable'', since it specifies the choice
of a particular hypersurface $y^{\mu}=y^{\mu}(x^{i})$.

In view of their parabolic nature, equations (\ref{m}) and (\ref{m1}) have a
space of solutions that, by an heuristic procedure, can be turned into an
Hilbert space, the inner product of which reads%
\begin{equation}
\langle\Psi_{1}\mid\Psi_{2}\rangle\equiv\int_{y^{\mu}=y^{\mu}(x^{i})}\Psi
_{1}^{\ast}\Psi_{2}D\phi\,\quad\frac{\delta\langle\Psi_{1}\mid\Psi_{2}\rangle
}{\delta y^{\mu}}=0,\label{n}%
\end{equation}
where $\Psi_{1}$ and $\Psi_{2}$ denote two generic solutions and $D\phi$ the
Lebesgue measure defined on the $\phi$-function space. The above inner product
induces the conserved functional probability distribution $\varrho
\equiv\langle\Psi\mid\Psi\rangle$.\newline The semiclassical limit of this
equations (\ref{m}) and (\ref{m1}) is obtained when taking $\hbar\rightarrow0$
and, by setting the wave functional as%
\begin{equation}
\Psi=expi\left\{  \frac{1}{\hbar}\Sigma(y^{\mu},\phi)\right\}  \label{o}%
\end{equation}
and then expanding $\Sigma$ in powers of $\hbar/i$, i. e.%
\begin{equation}
\Sigma=\Sigma_{0}+\frac{\hbar}{i}\Sigma_{1}+\left(  \frac{\hbar}{i}\right)
^{2}\Sigma_{2}+...\label{p}%
\end{equation}
By substituting (\ref{o}) and (\ref{p}) in equations (\ref{m}) and (\ref{m1}),
up to the zero-order approximation, we find the Hamilton-Jacobi equations%
\begin{equation}
-n^{\mu}\frac{\delta\Sigma_{0}}{\delta y^{\mu}}=\frac{1}{2\sqrt{h}}\left(
\frac{\delta\Sigma_{0}}{\delta\phi}\right)  ^{2}+\sqrt{h}\left(  \frac{1}%
{2}h^{ij}\partial_{i}\phi\partial_{j}\phi+V(\phi)\right)  ,\,\quad e_{i}^{\mu
}\frac{\delta\Sigma_{0}}{\delta y^{\mu}}=-\partial_{i}\phi\frac{\delta
\Sigma_{0}}{\delta\phi},\label{r}%
\end{equation}
which lead to the identification $\Sigma_{0}\equiv S^{\phi k}$.

\bigskip

\subsection{Reformulation of Quantum Geometrodynamics}

\label{sec2 rq}

We start by observing that, within the framework of a functional approach, a
covariant quantization of the 4-metric field is equivalent to take the wave
amplitude $\Psi= \Psi(g_{\mu\nu}(x^{\rho}))$; in the WDE approach, by adopting
the ADM slicing of the space-time, the problem is restated in terms of the
following replacement%
\begin{equation}
\Psi(g_{\mu\nu}(x^{\rho}))\rightarrow\Psi(N(t,x^{l}),N^{i}(t,x^{l}%
),h_{ij}(t,x^{l}))\,.\label{sss}%
\end{equation}
Then, since the lapse function $N$ and the shift vector $N^{i}$ are cyclic
variables, i.e. their conjugate momenta $p_{N}$ and $p_{N^{i}}$ vanish
identically, we get, on a quantum level, the following restrictions:%
\begin{equation}
\pi=0,\;\pi_{k}=0\,\quad\Rightarrow\,\quad\frac{\delta\Psi}{\delta
N}=0,\;\frac{\delta\Psi}{\delta N^{i}}=0\,;\label{sss1}%
\end{equation}
by other words, the wave functional $\Psi$ should be independent of $N$ and
$N^{i}$. Finally, the super-momentum constraint leads to the dependence of
$\Psi$ on the 3-geometries $\{h_{ij}\}$ (instead on a single 3-metric tensor
$h_{ij}$).\newline The criticism to the WDE approach, developed at the point
iii) of section \ref{sec1 WDE}, concerns with the ill-defined nature of the
replacement (\ref{sss}). The content of this section is entirely devoted to
reformulate the quantum geometrodynamics, by preserving the
(3+1)-representation of the space-time, but avoiding the ambiguity above
outlined in the WDE approach.

As outlined in the introduction to this section, we claim that the canonical
quantization of gravity has sense only when referred to a fixed slicing, in
which the notion of space or time like character of a vector field be
physically distinguishable.\newline To this aim we fix the lapse function and
the shift vector (now the slicing is fixed) and then we reparametrize the
gravitational action using the kinematical term (as in the assigned background
field theory), obtaining the total action:
\begin{align}
S^{g\phi k} &  =\int\limits_{\Sigma^{3}\times\Re}dtd^{3}x\left\{
p^{ij}\partial_{t}h_{ij}+\pi\partial_{t}N+\pi_{k}\partial_{t}N^{k}+p_{\phi
}\partial_{t}\phi+p_{\mu}\partial_{t}y^{\mu}\right.  +\nonumber\\
&  -\left.  \left(  \lambda\pi+\lambda^{i}\pi_{i}+N\left(  H^{g}+H^{\phi
}+H^{k}\right)  +N^{i}\left(  H_{i}^{g}+H_{i}^{\phi}+H_{i}^{k}\right)
\right)  \right\}  .\label{azione totale}%
\end{align}
Now the lapse function $N$ and the shift vector $N^{i}$ are to be again
regarded as dynamical variables (the slicing remain fixed by the hypersurfaces
parametric equations $y^{\mu}=y^{\mu}(t,x^{i})$ and by the vector $n^{\mu}$);
the new hamiltonian constraints are
\begin{align}
\pi &  =0,\qquad\pi_{k}=0,\label{new ham con}\\
H^{g}+H^{\phi}+H^{k} &  =0,\qquad H_{i}^{g}+H_{i}^{\phi}+H_{i}^{k}%
=0.\label{new ham con2}%
\end{align}
We note that the variation with respect to the dynamical field $y^{\mu}%
=y^{\mu}\left(  t,x^{i}\right)  $ and its conjugate momentum $p_{\mu}=p_{\mu
}\left(  t,x^{i}\right)  $ leads to the kinematical equation (\ref{eq din}).

Though from a mathematical point of view, to fix the reference frame is, in
view of the reparameterization which restores the canonical constraints, a
well defined procedure, it requires a physical interpretation; indeed the open
question is: which are the physical consequences of fixing the
slicing?\newline The complete answer to this question will be clear at the end
of this section, but now we can say that fixing the reference frame we modify
the physical system: the dynamical equations and the constraints, describe no
more the dynamics of the initial system formed by gravitational and scalar
field, but the addition of the kinematical term introduces a new physical
field, which, as shown in section \ref{sec2 ka} can be interpreted as a dust
interacting with a gravito-electromagnetic-like field. We remark that in a
purely classical system it is not necessary to introduce this additional term
to the gravity-matter action and therefore we expect that the dust has
effects on the dynamics of those systems which evolve from a quantum state.

Now to quantize the new constraints (\ref{new ham con}), (\ref{new ham con2})
we use the canonical procedure, by implementing the canonical variables to
quantum operators. We assume that the state of the gravitational and matter
system be described by a wave functional $\Psi=\Psi\left(  y^{\mu},\phi
,h_{ij},N,N^{i}\right)  $. Then the new quantum dynamics of the whole system
is now described by the functional differential system:%
\begin{align}
\dfrac{\delta\Psi}{\delta N} &  =0,\qquad\qquad\qquad\qquad\qquad\dfrac
{\delta\Psi}{\delta N^{i}}=0,\\
i\hbar n^{\mu}\dfrac{\delta\Psi}{\delta y^{\mu}}= &  \left(  \widehat{H}%
^{g}+\widehat{H}^{\phi}\right)  \Psi,\qquad i\hbar\partial_{i}y^{\mu}%
\dfrac{\delta\Psi}{\delta y^{\mu}}=\left(  \widehat{H}_{i}^{g}+\widehat{H}%
_{i}^{\phi}\right)  \Psi,\label{eq quantistiche}%
\end{align}
being $\widehat{H}^{g}+\widehat{H}^{\phi}$and $\widehat{H}_{i}^{g}+\widehat
{H}_{i}^{\phi}$ the hamiltonian operators after the quantum implementation of
the canonical variables. By the first line equations, the wave functional does
not depend on the lapse function $N$ and the shift vector $N^{i},$ so, since
now, we limit our attention on the other two equations, considering that the
wave functional $\Psi$ depends only on the 3-metric $h_{ij}\left(
x^{k}\right)  $, the scalar field $\phi\left(  x^{k}\right)  $ and the new
field $y^{\mu}\left(  x^{k}\right)  ,$ which plays the role of a time
variable, by specifying the hypersurface on which the wave functional is taken
(we stress how its spatial gradients behaves like potential terms).

Moreover, the second of equation (\ref{eq quantistiche}) still assures the
invariance of the wave functional under the spatial diffeomorphism. Then,
denoting by $\left\{  h_{ij}\right\}  $ a whole class of 3-geometries (i.e.
connected via 3-coordinates reparameterization), the wave functional should
yet be taken on such more appropriate variable instead of a special
realization of the 3-metric.\newline In the first of equations (\ref{eq
quantistiche}) the vector field $n^{\mu}\left(  y^{\rho}\right)  $ is an
arbitrary one, without any peculiar geometrical meaning; but when taking into
account the first of kinematical equation (\ref{eq din}), $n^{\mu}$ becomes a
real unit normal vector field, since, once fixed $N$ and $N^{i},$ $y^{\mu
}\left(  t,x^{i}\right)  $ pays the price for its geometrical interpretation.
These considerations lead us to claim that the first of equation (\ref{eq
din}) should be included in the dynamics even on the quantum level. The
physical justification for this statement relies on the fact that no
information about the dynamic of the kinematical dust comes from such an
equation has discussed in the previous section; in fact there we have shown
how the whole dynamics of the dust be entirely contained in the momentum
equation. In agreement to what we said in the introduction to this work, the
surviving of this classical equation on a quantum level, reflects the
classical nature of the ``device'' operating the (3 + 1)-splitting.\newline To
take into account this equation is equivalent to reduce $y^{\mu}$ to a simple
$\infty-$dimensional parameter for the system dynamics.

In agreement with this point of view, we can smear the quantum dynamics on a
whole 1-parameter family of spatial hypersurfaces $\Sigma_{t}^{3}$ filling the
space-time; as soon as we introduce the notation
\begin{equation}
\partial_{t}=\underset{\Sigma_{t}^{3}}{\int}d^{3}x\partial_{t}y^{\mu}%
\dfrac{\delta}{\delta y^{\mu}},
\end{equation}
then equations (\ref{eq quantistiche}) acquire the Schr\"{o}dinger form
\begin{equation}
i\hbar\partial_{t}\Psi=\widehat{\mathcal{H}}\Psi,\label{raffaella3}%
\end{equation}
where%
\begin{equation}
\widehat{\mathcal{H}}=\underset{\Sigma_{t}^{3}}{\int}d^{3}x\left[  N\left(
\widehat{H}^{g}+\widehat{H}^{\phi}\right)  +N^{i}\left(  \widehat{H}_{i}%
^{g}+\widehat{H}_{i}^{\phi}\right)  \right]  .\label{operatore hamiltoniano}%
\end{equation}
In this new framework the wave functional can be taken directly on the label
time $\left(  \text{i.e. }\Psi=\Psi\left(  t,\phi,h_{ij}\right)  \right)  $
(where we have removed the curl bracket from $h_{ij}$ because, now, the wave functional is no longer invariant under
3-diffeomorphism), since the latter becomes a physical clock via the
correspondence, we show below, between the eigenvalue problem of the equation
(\ref{raffaella3}) and the energy-momentum of the dust discussed in the
previous section.

In order to construct the Hilbert space associated to the Schr\"{o}dinger-like
equation we must prove the hermitianity of the hamiltonian operator; since the
hermitian character of the $\phi$ term was proved in \cite{Kuc1981}, as well as of the operator $\widehat{H}^{g}$ in \cite{Mon2002} under the
following choice for the normal ordering
\begin{equation}
G_{ijkl}p^{ij}p^{kl}\rightarrow-\hbar^{2}\dfrac{\delta}{\delta h_{ij}}\left(
G_{ijkl}\dfrac{\delta\left(  ...\right)  }{\delta h_{kl}}\right)
,\label{ordinamento}%
\end{equation}
then it remains to be shown the hermitian character of the operator
$\widehat{h}=\underset{\Sigma_{t}^{3}}{\int}d^{3}xN^{i}\widehat{H}_{i}^{g}$.
In Dirac notation we have to show that:
\begin{equation}
\left\langle \Psi_{1}\left|  \widehat{h}\right|  \Psi_{2}\right\rangle
=\left\langle \Psi_{2}\left|  \widehat{h}\right|  \Psi_{1}\right\rangle
^{\ast}.
\end{equation}
To this aim we write down the explicit expression of the above bracket:
\begin{equation}
\left\langle \Psi_{1}\left|  \widehat{h}\right|  \Psi_{2}\right\rangle
=2i\hbar{\displaystyle\int\limits_{\mathcal{F}_{t}}}Dh\underset{\Sigma_{t}%
^{3}}{\int}d^{3}x\Psi_{1}^{\ast}N^{i}h_{ik}\nabla_{j}\dfrac{\delta}{\delta
h_{kj}}\Psi_{2},\label{braket}%
\end{equation}
where $Dh$ is the Lebesgue measure in the 3-geometries functional space.Now
integrating by parts, considering that the hypersurfaces $\Sigma_{t}^{3}$ are
compact and using, in view of the functional Gauss theorem, the following
relation:%
\begin{equation}
{\displaystyle\int\limits_{\mathcal{F}_{t}}}Dh\underset{\Sigma_{t}^{3}%
}{{\displaystyle\int}}d^{3}x\dfrac{\delta}{\delta h_{kj}}\left(
......\right)  =0,\label{gauss}%
\end{equation}
we can rewrite the expression (\ref{braket}) in the following form:%
\begin{equation}
\left\langle \Psi_{1}\left|  \widehat{h}\right|  \Psi_{2}\right\rangle
=2i\hbar{\displaystyle\int\limits_{\mathcal{F}_{t}}}Dh\underset{\Sigma_{t}%
^{3}}{{\displaystyle\int}}d^{3}x\dfrac{\delta}{\delta h_{kj}}\left(  \Psi
_{1}^{\ast}\left(  \nabla_{j}N^{i}\right)  h_{ik}\right)  \Psi_{2}%
.\label{integr}%
\end{equation}
It is possible to show that two of the terms, which come from the right side
of (\ref{integr}) when the functional derivative operates on the quantities in
the parenthesis, are zero. In fact, acting with the functional derivative on
the 3-metric, we obtain:%
\begin{equation}
2i\hbar{\displaystyle\int\limits_{\mathcal{F}_{t}}}Dh\underset{\Sigma_{t}^{3}%
}{{\displaystyle\int}}d^{3}x\Psi_{1}^{\ast}\left(  \nabla_{j}N^{i}\right)
\dfrac{\delta h_{ik}}{\delta h_{kj}}\Psi_{2}=-2i\hbar%
%TCIMACRO{\dint \limits_{\mathcal{F}_{t}}}%
%BeginExpansion
{\displaystyle\int\limits_{\mathcal{F}_{t}}}
%EndExpansion
Dh\underset{\Sigma_{t}^{3}}{%
%TCIMACRO{\dint }%
%BeginExpansion
{\displaystyle\int}
%EndExpansion
}d^{3}x\nabla_{j}\left(  \Psi_{1}^{\ast}\Psi_{2}\right)  N^{j},\label{zero}%
\end{equation}
where we have integrated by parts and used the compactness of the
hypersurfaces $\Sigma_{t}^{3}.$ But the right hand side of (\ref{zero}) is
zero, because $\Psi$ is a functional, so it does not depend on $x.$%
\newline When the functional derivative in expression (\ref{integr}) acts on
the covariant derivative of the shift vector, we obtain:%
\begin{align}
2i\hbar%
%TCIMACRO{\dint \limits_{\mathcal{F}_{t}}}%
%BeginExpansion
{\displaystyle\int\limits_{\mathcal{F}_{t}}}
%EndExpansion
Dh\underset{\Sigma_{t}^{3}}{%
%TCIMACRO{\dint }%
%BeginExpansion
{\displaystyle\int}
%EndExpansion
} &  d^{3}xh_{ik}\Psi_{1}^{\ast}\Psi_{2}\dfrac{\delta}{\delta h_{kj}}\left(
\nabla_{j}N^{i}\right)  =\nonumber\\
&  =2i\hbar%
%TCIMACRO{\dint \limits_{\mathcal{F}_{t}}}%
%BeginExpansion
{\displaystyle\int\limits_{\mathcal{F}_{t}}}
%EndExpansion
Dh\underset{\Sigma_{t}^{3}}{%
%TCIMACRO{\dint }%
%BeginExpansion
{\displaystyle\int}
%EndExpansion
}d^{3}xh_{ik}\Psi_{1}^{\ast}\Psi_{2}\dfrac{\delta}{\delta h_{kj}}\left(
\Gamma_{jm}^{i}N^{m}\right)  ,\label{zero2}%
\end{align}
since in the right side term, the derivative operator is applied to a function
of $x$ and not to a functional, thus, like in the case of the variation with
respect a dynamical variable, the ordinary derivative operator and the
functional one commute, so it is simple to show that $\dfrac{\delta}{\delta
h_{kj}}\left(  \Gamma_{jm}^{i}N^{m}\right)  =0,$ thus the term (\ref{zero2})
is identically zero.\newline Finally the expression (\ref{integr}) can be
rewrite:%
\begin{align}
\left\langle \Psi_{1}\left|  \widehat{h}\right|  \Psi_{2}\right\rangle  &
=2i\hbar%
%TCIMACRO{\dint \limits_{\mathcal{F}_{t}}}%
%BeginExpansion
{\displaystyle\int\limits_{\mathcal{F}_{t}}}
%EndExpansion
Dh\underset{\Sigma_{t}^{3}}{%
%TCIMACRO{\dint }%
%BeginExpansion
{\displaystyle\int}
%EndExpansion
}d^{3}x\dfrac{\delta\Psi_{1}^{\ast}}{\delta h_{kj}}\left(  \nabla_{j}%
N^{i}\right)  h_{ik}\Psi_{2}=\nonumber\\
&  =-2i\hbar%
%TCIMACRO{\dint \limits_{\mathcal{F}_{t}}}%
%BeginExpansion
{\displaystyle\int\limits_{\mathcal{F}_{t}}}
%EndExpansion
Dh\underset{\Sigma_{t}^{3}}{%
%TCIMACRO{\dint }%
%BeginExpansion
{\displaystyle\int}
%EndExpansion
}d^{3}x\Psi_{2}N^{i}h_{ik}\nabla_{j}\dfrac{\delta\Psi_{1}^{\ast}}{\delta
h_{kj}}=\left\langle \Psi_{2}\left|  \widehat{h}\right|  \Psi_{1}\right\rangle
^{\ast}.
\end{align}
The above equality assures $\widehat{\mathcal{H}}$ is an Hermitian operator.
Defining the following inner product:
\begin{equation}
\left\langle \Psi_{1}\mid\Psi_{2}\right\rangle =\underset{y_{t}}{\int}%
DhD\phi\Psi_{1}^{\ast}\Psi_{2},\label{bracket}%
\end{equation}
where $DhD\phi$ is the Lebesgue measure for the functional space of all the
dynamical variables and $y_{t}$ is the corresponding functional domain, we can
turn the space of solutions of the Schr\"{o}dinger-like equation into an
Hilbert space. We interpret the above bracket as the probability that a state
$\left|  \Psi_{1}\right\rangle $ falls into another state $\left|  \Psi
_{2}\right\rangle $ and, defining the density of probability $\rho=\Psi^{\ast
}\Psi,$ we can also construct the amplitude for the system lying in a field
configuration. By the hermitian character of the operator $\widehat
{\mathcal{H}},$ it is possible to show that the probability is constant in
time, in fact:
\begin{equation}
\partial_{t}\left\langle \Psi_{1}\right|  \left.  \Psi_{2}\right\rangle
=\underset{\Sigma_{t}^{3}}{%
%TCIMACRO{\dint }%
%BeginExpansion
{\displaystyle\int}
%EndExpansion
}d^{3}x\partial_{t}y^{\mu}\dfrac{\delta}{\delta y^{\mu}}\left\langle \Psi
_{1}\right|  \left.  \Psi_{2}\right\rangle =\dfrac{i}{\hbar}\left(
\left\langle \widehat{\mathcal{H}}\Psi_{1}\mid\Psi_{2}\right\rangle
-\left\langle \Psi_{1}\mid\widehat{\mathcal{H}}\Psi_{2}\right\rangle \right)
=0,
\end{equation}
the general character of the deformation vector allows us to write the
fundamental conservation law
\begin{equation}
\dfrac{\delta\left\langle \Psi_{1}\right|  \left.  \Psi_{2}\right\rangle
}{\delta y^{\mu}}=0,
\end{equation}
which assures the probability does not depend on the choice of the hypersurface.

The density of probability $\rho$ satisfies a continuity equation, which can
be obtained multiplying the Schr\"{o}dinger-like equation times the complex
conjugate wave function $\Psi^{\ast}$ and the complex conjugate equation times
the wave function $\Psi,$ i.e.
\begin{equation}
i\hbar\Psi^{\ast}\partial_{t}\Psi=\Psi^{\ast}\widehat{\mathcal{H}}\Psi
,\qquad-i\hbar\Psi\partial_{t}\Psi^{\ast}=\Psi\widehat{\mathcal{H}}^{\ast}%
\Psi^{\ast},\label{equazioni sopra}%
\end{equation}
subtracting the second of equation (\ref{equazioni sopra}) from the first one,
we obtain:%
\begin{align}
i\hbar\partial_{t}\left(  \Psi\Psi^{\ast}\right)   &  =\underset{\Sigma
_{t}^{3}}{%
%TCIMACRO{\dint }%
%BeginExpansion
{\displaystyle\int}
%EndExpansion
}d^{3}x\left\{  -\hbar^{2}\left(  \Psi^{\ast}N\dfrac{\delta}{\delta h_{ij}%
}G_{ijkl}\dfrac{\delta}{\delta h_{kl}}\Psi-\Psi N\dfrac{\delta}{\delta h_{ij}%
}G_{ijkl}\dfrac{\delta}{\delta h_{kl}}\Psi^{\ast}\right)  \right.
+\nonumber\\
&  -\hbar^{2}\left(  \Psi^{\ast}\dfrac{N}{2\sqrt{h}}\dfrac{\delta}{\delta\phi
}\dfrac{\delta}{\delta\phi}\Psi-\Psi\dfrac{N}{2\sqrt{h}}\dfrac{\delta}%
{\delta\phi}\dfrac{\delta}{\delta\phi}\Psi^{\ast}\right)  +\nonumber\\
&  +2i\hbar\left(  \Psi^{\ast}N^{i}h_{ik}\nabla_{j}\dfrac{\delta}{\delta
h_{kj}}\Psi+\Psi N^{i}h_{ik}\nabla_{j}\dfrac{\delta}{\delta h_{kj}}\Psi^{\ast
}\right)  +\nonumber\\
&  -\left.  i\hbar\left(  \Psi^{\ast}N^{i}\partial_{i}\phi\dfrac{\delta
}{\delta\phi}\Psi+\Psi N^{i}\partial_{i}\phi\dfrac{\delta}{\delta\phi}%
\Psi^{\ast}\right)  \right\}  ,\label{eq cont}%
\end{align}
defining now the tensor probability current $A_{ij}$, which is connected with
the 3-metric tensor field, in the following way:%
\begin{equation}
A_{ij}=-i\hbar\left(  \Psi^{\ast}NG_{ijkl}\dfrac{\delta}{\delta h_{kl}}%
\Psi-\Psi NG_{ijkl}\dfrac{\delta}{\delta h_{kl}}\Psi^{\ast}\right)
+2h_{ki}\left(  \nabla_{j}N^{k}\right)  \Psi^{\ast}\Psi,
\end{equation}
and the scalar probability current $A,$ connected, instead, to the presence of
the scalar field $\phi,$ as:%
\begin{equation}
A=-i\hbar\dfrac{N}{2\sqrt{h}}\left(  \Psi^{\ast}\dfrac{\delta}{\delta\phi}%
\Psi-\Psi\dfrac{\delta}{\delta\phi}\Psi^{\ast}\right)  -i\hbar\left(
\phi\partial_{i}N^{i}\Psi^{\ast}\Psi\right)  ,
\end{equation}
the equation (\ref{eq cont}) takes the following form:
\begin{equation}
\partial_{t}\rho+\underset{\Sigma_{t}^{3}}{%
%TCIMACRO{\dint }%
%BeginExpansion
{\displaystyle\int}
%EndExpansion
}d^{3}x\left(  \dfrac{\delta A_{ij}}{\delta h_{ij}}+\dfrac{\delta A}%
{\delta\phi}\right)  =0,
\end{equation}
integrating on the functional space $y_{t}$, using the generalized Gauss
theorem (\ref{gauss}), the continuity equation assures that the probability is
constant in time as above.

Let us now reconsider the Schr\"{o}dinger dynamics in terms of a time
independent eigenvalues problem. To this end we expand the wave functional as
follows:
\begin{align}
\Psi\left(  t,\phi,h_{ij}\right)  =\underset{y_{t}^{\ast}}{%
%TCIMACRO{\dint }%
%BeginExpansion
{\displaystyle\int}
%EndExpansion
}D\Omega &  DK\Theta\left(  \Omega,K_{i}\right)  \chi_{\Omega,K_{i}}\left(
\phi,h_{ij}\right)  \cdot\nonumber\\
&  \cdot\exp\left\{  -\dfrac{i}{\hbar}\int\limits_{t_{0}}^{t}dt^{\prime
}\underset{\Sigma_{t}^{3}}{\int}d^{3}x\left(  N\Omega+N^{i}K_{i}\right)
\right\}  ,\label{expan}%
\end{align}
being $t_{0}$ an assigned initial ``instant''. Where $D\Omega DK$ denotes the
Lebesgue measure in the functional space $y_{t}^{\ast}$ of the conjugate
function $\Omega\left(  x^{i}\right)  $ and $K_{i}\left(  x^{i}\right)  ,$
$\Theta=\Theta\left(  \Omega,K_{i}\right)  $ a functional valued in this
domain, whose form is determined by the initial conditions $\Psi_{0}%
=\Psi\left(  t_{0},\phi,h_{ij}\right)  $. When we substitute the expansion (\ref{expan}) of the wave functional into 
(\ref{raffaella3}), such equation is satisfied only if the following 
$\infty^{3}-$dimensional eigenvalues problem takes place:%
\begin{equation}
\left(  \widehat{H}^{g}+\widehat{H}^{\phi}\right)  \chi_{_{\Omega,K_{i}}%
}=\Omega\left(  x^{j}\right)  \chi_{_{\Omega,K_{i}}},\qquad\left(  \widehat
{H}_{i}^{g}+\widehat{H}_{i}^{\phi}\right)  \chi_{_{\Omega,K_{i}}}=K_{i}\left(
x^{j}\right)  \chi_{_{\Omega,K_{i}}}.\label{eigenvalue}%
\end{equation}
Now to characterize the physical meaning of the above eigenvalues, we
construct the semi-classical limit of the Schr\"{o}dinger-like equation, by
splitting the wave functional into its modulus and phase, as follows:
\begin{equation}
\Psi=\sqrt{\rho}e^{\tfrac{i}{\hbar}\sigma}.\label{psi semiclassical}%
\end{equation}
Then in the limit $\hbar\rightarrow0$ we obtain for $\sigma$ an
Hamilton-Jacobi equation of the form:
\begin{align}
-\partial_{t}\sigma &  =\underset{\Sigma_{t}^{3}}{%
%TCIMACRO{\dint }%
%BeginExpansion
{\displaystyle\int}
%EndExpansion
}d^{3}xN\left(  G_{ijkl}\dfrac{\delta\sigma}{\delta h_{ij}}\dfrac{\delta
\sigma}{\delta h_{ij}}-\sqrt{h}^{\left(  3\right)  }R\right.  +\nonumber\\
&  +\left.  \dfrac{1}{2\sqrt{h}}\dfrac{\delta\sigma}{\delta\phi}\dfrac
{\delta\sigma}{\delta\phi}+\dfrac{\sqrt{h}}{2}h^{ij}\partial_{i}\phi
\partial_{j}\phi+\sqrt{h}V\left(  \phi\right)  \right)  +\nonumber\\
&  -\underset{\Sigma_{t}^{3}}{%
%TCIMACRO{\dint }%
%BeginExpansion
{\displaystyle\int}
%EndExpansion
}d^{3}xN^{i}\left(  2h_{ik}\nabla_{j}\dfrac{\delta\sigma}{\delta h_{kj}%
}-\partial_{i}\phi\dfrac{\delta\sigma}{\delta\phi}\right)  \label{ham-jac}%
\end{align}
The non vanishing of the $\sigma$ time derivative reflects the evolutive
character appearing in the constructed theory and makes account for the
presence, on the classical limit, of the dust matter discussed in the previous
section. To clarify this feature, we set
\begin{equation}
\sigma\left(  t,\phi,h_{ij}\right)  =\tau\left(  \phi,h_{ij}\right)
+\int\limits_{t_{0}}^{t}dt^{\prime}\underset{\Sigma_{t}^{3}}{%
%TCIMACRO{\dint }%
%BeginExpansion
{\displaystyle\int}
%EndExpansion
}d^{3}x\left(  N\Omega+N^{i}K_{i}\right)  .
\end{equation}
When we substitute this expression in the Hamilton-Jacobi equation, and
identify $p^{ij}=\dfrac{\delta\tau}{\delta h_{ij}},$ $p_{\phi}=\dfrac
{\delta\tau}{\delta\phi},$ then the equation (\ref{ham-jac}) becomes
equivalent to the $\infty-$dimensional ones:%
\begin{equation}
\left(  H^{g}+H^{\phi}\right)  =\Omega\left(  x^{j}\right)  \qquad\left(
H_{i}^{g}+H_{i}^{\phi}\right)  =K_{i}\left(  x^{j}\right)
\end{equation}
We stress how these equations coincides with those ones obtainable by the
eigenvalues problem (\ref{eigenvalue}), as soon as we choose the classical
limit of $\chi\sim e^{\tfrac{i}{\hbar}\tau}$ thus, at the end of this
analysis, recalling expressions (\ref{raffaella}) and (\ref{new ham con2}), we
can identify the super-hamiltonian eigenvalue $\Omega$ with $\omega$ and the
super-momentum eigenvalues $K_{i}$ with $k_{i}.$ On the other hand by
equations (\ref{raffaella}) and (\ref{ham mom2}), the above identification
implies that: $\Omega=-\sqrt{h}\varepsilon$ and $K_{i}=-\gamma A_{i}\omega.$

The relation we obtained show how super-hamiltonian and super-momentum
eigenvalues are directly connected with the dust fields introduced in section
\ref{sec2 ka}. Even starting from a quantum point of view we recognize the
existence of a dust fluid playing the role of a physical clock for the
gravity-matter dynamics.

\bigskip

\section{A Simple Cosmological Model}

\label{sec3}

If the theory here proposed is a predictive one, we should expect to observe
the trace of this reference fluid energy density from all those systems which
underwent a classical limit; such a situation is surely true for our actual
Universe and, indeed, we really observe (in the synchronous reference of our
galaxy) an unidentified dust energy, the so-called \emph{dark matter}; in the
next two sub-sections, we will try to understand if it can exist a correlation
between our dust fluid and the observed ``matter component'' of the Universe.

\bigskip

\subsection{3-Diffeomorphisms Invariant Theory}

\label{sec3 3d}

Before to discuss the application of our theory to a FRW Universe, we want to
rewrite the above reformulation of quantum geometrodynamics preserving the
3-diffeomorphisms invariance. This means that the quantum equation take the
following form:%
\begin{equation}
i\hbar n^{\mu}\frac{\delta\Psi}{\delta y^{\mu}}=(\widehat{H}^{g}+\widehat
{H}_{i}^{\phi})\Psi\,,\quad(\widehat{H}_{i}^{g}+\widehat{H}_{i}^{\phi}%
)\Psi=0\,,\quad\Psi=\Psi(\{h_{ij}\},y^{\mu}),\label{a1x}%
\end{equation}
where now the wave functional is taken again on the 3-geometries ($\{h_{ij}%
\}$) related by the 3-diffeomorphisms.\newline These ($4\times\infty^{3}$)
equations, which correspond to a natural extension of the Wheeler-De Witt
approach, have the fundamental feature that again the first of them is
parabolic and it is just this property which still allows to overcome the
limits of the WDE above discussed.Though this set of equations provides a
satisfactory description of the 3-geometries quantum dynamics, nevertheless it
turns out convenient and physically meaningful to take, by (\ref{eq din}), the
wave functional evolution along a one-parameter family of spatial
hypersurfaces, filling the Universe.

By the first of equations (\ref{eq din}), the above (\ref{a1x}) can be
rewritten as follows:%
\begin{equation}
i\hbar\frac{\delta\Psi}{\delta y^{\mu}}\partial_{t}y^{\mu}=N(\widehat{H}%
^{g}+\widehat{H}^{\phi})\Psi.\label{b1}%
\end{equation}
Now this set of equations can be (heuristically) rewritten as a single one by
integrating over the hypersurfaces $\Sigma_{t}^{3}$, i. e.%
\begin{equation}
i\hbar\partial_{t}\Psi=i\hbar\int_{\Sigma_{t}^{3}}\left\{  \frac{\delta\Psi
}{\delta y^{\mu}}\partial_{t}y^{\mu}\right\}  d^{3}x=\hat{\mathcal{H}}%
\Psi\equiv\left[  \int_{\Sigma_{t}^{3}}N(\hat{H}^{g}+\hat{H}^{\phi}%
)d^{3}x\right]  \Psi.\label{c1}%
\end{equation}
The above equations (\ref{c1}) and (\ref{a1x}) show how in the present
approach the wave functional is still no longer invariant under infinitesimal
displacements of the time variable.

It is possible to show that, like above, the operator $\widehat{\mathcal{H}}$
is an hermitian one, so we still have the fundamental conservation law%
\begin{equation}
\frac{\delta\langle\Psi_{1}\mid\Psi_{2}\rangle}{\delta y^{\mu}}=0. \label{o99}%
\end{equation}
Substituting the usual expansion%
\begin{equation}
\Psi(y^{\mu},\{h_{ij}\},\phi)=\int_{{}^{\ast}\mathcal{Y}_{t}}D\omega
\Theta(\omega)\chi_{\omega}(\{h_{ij}\},\phi)exp\left\{  \frac{i}{\hbar}%
\int_{\Sigma_{t}^{3}}d^{3}x\int dy^{\mu}(\omega n_{\mu})\right\}  \label{q1}%
\end{equation}
into equations (\ref{a1x}) we get the eigenvalues problems%
\begin{equation}
(\hat{H}^{g}+\hat{H^{\phi}})\chi_{\omega}=\omega\chi_{\omega}\,\quad
(\hat{H}_{i}^{g}+\hat{H}_{i}^{\phi})\chi_{\omega}=0\label{r1}%
\end{equation}
Here $\omega(x^{i})$ is not a 3-scalar, but it transforms, under
3-diffeomorphisms, like $\hat{H}^{g}$ or $\hat{H}^{\phi}$, so ensuring that
$\omega d^{3}x$, as it should, be an invariant quantity.\newline Now we
observe that, by (\ref{eq din}), equation (\ref{q1}) rewrites%
\begin{align}
\Psi(y^{\mu},\{h_{ij}\},\phi)  & =\int_{{}^{\ast}\mathcal{Y}_{t}}D\omega
\Theta(\omega)\chi_{\omega}(\{h_{ij}\},\phi)exp\left\{  \frac{i}{\hbar}%
\int_{\Sigma_{t}^{3}}d^{3}x\int_{t_{0}}^{t}dt^{\prime}\partial_{t^{\prime}%
}y^{\mu}(\omega n_{\mu})\right\}  =\nonumber\\
& =\int_{{}^{\ast}\mathcal{Y}_{t}}D\omega\Theta(\omega)\chi_{\omega}%
(\{h_{ij}\},\phi)exp\left\{  -\frac{i}{\hbar}\int_{t_{0}}^{t}dt^{\prime}%
\int_{\Sigma_{t}^{3}}d^{3}x(N\omega)\right\}  ,\label{u1}%
\end{align}
being $t_{0}$ an assigned initial ``instant.`\newline To the same result we
could arrive by choosing, without any loss of generality, the coordinates
system ($t,x^{i}$), i. e. $y^{0}\equiv t\;,y^{i}\equiv x^{i}$; indeed, for
this system, the spatial hypersurfaces have equation $t=const,$ i. e.
$dy^{\mu}\rightarrow(dt,0,0,0)$ and we have $n_{0}=N$. By other words the wave
functional (\ref{u1}) is to be interpreted directly in terms of the time
variable $t$, i. e. $\Psi(\{h_{ij}\},\phi,t)$ and, in fact, it turns out
solution of the wave equation%
\begin{equation}
i\hbar\partial_{t}\Psi(\{h_{ij}\},\phi,t)=\hat{\mathcal{H}}\Psi(\{h_{ij}%
\},\phi,t)\label{v1}%
\end{equation}
The expansion (\ref{u1}) of the wave functional and the eigenvalues problems
(\ref{r1}) completely describe the quantum dynamics of the 3-geometries.

In this 3-diffeomorphisms invariant approach it is very simple to show that
the fluid of reference reduces to a real dust with the energy momentum tensor
\begin{equation}
T^{\mu\nu}=\varepsilon n^{\mu}n^{\nu}.
\end{equation}
To conclude, it is worth remarking how, the main difference between our
approach and others interesting ones, that lead to the same formal issue (see
the discussion in the appendix about the comparison with the so-called
``multi-time approach'' as well as the formulations presented in
\cite{Rov1991a, Rov1991b} and \cite{Smo1993, RovSmo1993}), consists of, in
the latter, the super-hamiltonian is preliminary reduced to a linear form,
and, overall, of setting \emph{ad hoc} fields which play the role of time (for
instance in \cite{Smo1993,RovSmo1993} is postulated, in the theory, the
presence of a real mass-less scalar field), in the former, in stead, we simply
extend to the 3-metric dynamics the kinematical (embedding-like) action to
provide physical meaning in the splitting procedure, and then interpret it as
a dust fluid (with the role of time). In this scheme the 3-metric is related
to the space-time one by the dynamical field $y^{\mu}$, so, heuristically, we
can say to bypass the \emph{theory background independence}.

\bigskip

\subsection{FRW Quantum Universe}

\label{sec3 frw}

Since the clock by which we are measuring the age of the Universe is
(essentially) a synchronous one, and we expect the cosmological dynamics
became a classical one, then the contribution of the ``dust fluid'' energy
density must appear in the galaxies recession. Below we will face the
questions about the modifications introduced, by our approach, in the quantum
evolution of the Universe, and about the actual value of the dust energy density.

We investigate the quantum dynamics predicted, in a synchronous reference, by
equation (\ref{v}) for the closed Friedmann-Robertson-Walker model
\cite{KolTur1990,Har1988}, whose line element reads (below we adopt the
standard notations for the fundamental constants)%
\begin{equation}
ds^{2}=-c^{2}dt^{2}+R_{c}^{2}(t)[d\xi^{2}+\sin^{2}\xi(d\eta^{2}+\sin^{2}\eta
d\phi^{2})]\,,\label{yyy}%
\end{equation}
where $0\leq\xi<\pi,$ $0\leq\eta<\pi,$ $0\leq\phi<2\pi.$ Here $R_{c}$ denotes
the radius of curvature of the Universe, measurable, in principle, via the
relation $R_{c}=c/(H\sqrt{\bar{\Omega}-1})$ (being $H$ the Hubble function,
$\bar{\Omega}$ the critical parameter and $R_{c(today)}\sim\mathcal{O}%
(10^{28}cm)$). 

In the very early phases of the Universe evolution, it is expected a space
filled by a thermal bath, involving all the fundamental particles; since, at
very high temperatures, all the massive particles are ultra relativistic ones,
then the most appropriate phenomenological representation of the
matter-radiation thermal bath, is provided by an energy density of the form
$\mu^{2}/R_{c}^{4}$.\newline Furthermore, the idea that the Universe underwent
an inflationary scenario, leads us to include \emph{ab initio} in the dynamics
a real self-interacting scalar field $\phi$, described by a
``finite-temperature'' potential $V_{T}(\phi)$ (here $T$ denotes the thermal
bath temperature), which we may take, for instance, in the Coleman-Weinberg form%
\begin{equation}
V_{T}(\phi)=\frac{B\sigma^{4}}{2h^{3}c^{3}}+B\frac{\phi^{4}}{hc}\left[
\ln\left(  \frac{l_{Pl}\phi^{2}}{\sigma^{2}}\right)  -\frac{1}{2}\right]
+\frac{1}{2}{m_{T}}^{2}\phi^{2}\,\quad m_{T}=\sqrt{\lambda T^{2}-m^{2}%
}\,,\label{CW}%
\end{equation}
with $(m,\lambda)=const.,$ $B$ is a parameter related to the fundamental
constraints of the theory (estimated $\mathcal{O}(10^{-3})$, $\sigma$
corresponds to the energy scale associated with the symmetry breaking process
(i.e. $\sigma\sim\mathcal{O}(10^{15})GeV)$), while $m$ and $l_{Pl}$ denote,
respectively, the inverse of a characteristic length and $l_{Pl}$ the Planck
length $l_{Pl}\equiv\sqrt{G\hbar/c^{3}}$.; the temperature dependence of the
potential term can be also regarded as a time evolution of the model.\newline
The dynamics of such a cosmological model is summarized, as shown when
developing the Einstein-Hilbert action under the present symmetries, by the
hamiltonian function%
\begin{equation}
\frac{\mathcal{H}}{c}=-\frac{l_{Pl}^{2}}{3\pi\hbar}\frac{p_{R_{c}}^{2}}{R_{c}%
}+\frac{c}{4\pi^{2}}\frac{p_{\phi}^{2}}{R_{c}^{3}}+\frac{\mu^{2}}{R_{c}}%
-\frac{3\pi\hbar}{4l_{Pl}^{2}}R_{c}+2\pi^{2}R_{c}^{3}V_{T}(\phi)\,,\label{w}%
\end{equation}
with $p_{R_{c}}$ and $p_{\phi}$ being the conjugate momenta to $R_{c}$ and
$\phi$.

Thus, the Schr\"odinger equation (\ref{v}) reads, once turned the above
hamiltonian into an operator (which possesses the right normal ordering), as follows%
\begin{align}
& \frac{i\hbar}{c}\partial_{t}\Psi(t,R_{c},\phi)=\nonumber\\
& =\left\{  \frac{l_{Pl}^{2}\hbar}{3\pi}\partial_{R_{c}}\frac{1}{R_{c}%
}\partial_{R_{c}}-\frac{\hbar^{2}c}{4\pi^{2}}\frac{1}{R_{c}^{3}}\partial
_{\phi}^{2}+\frac{\mu^{2}}{R_{c}}-\frac{3\pi\hbar}{4l_{Pl}^{2}}R_{c}+2\pi
^{2}R_{c}^{3}V_{T}(\phi)\right\}  \Psi(t,R_{c},\phi)\,,\label{z}%
\end{align}
Before going on with the analysis of this equation, we need to precise some
aspects concerning the potential term relevance during the Universe
evolution.\newline It is well-known that the classical scalar field dynamics
is governed by the following equation%
\begin{equation}
\ddot{\phi}+3H\dot{\phi}+c^{2}\hbar^{2}\frac{dV_{T}}{d\phi}=0\,.\label{cdf}%
\end{equation}
The presence of the potential term is surely crucial to generate the
inflationary scenario, but, sufficiently close to the initial ``Big-Bang'',
its dynamical role is expected to be very limited; in fact, if we neglect the
potential term in (\ref{cdf}), then, remembering that for early times
$R_{c}\sim\sqrt{t}\;\rightarrow\;H\sim1/2t$, we get the free field solution
$\phi\propto\ln t$. Now the terms we retained to solve equation (\ref{cdf})
are potentially of the order $\mathcal{O}(1/t^{2})$; in the limit toward the
``Big-Bang'' ($t\rightarrow0$), the potential term (\ref{CW}) (we recall that
$T\propto1/R_{c}\propto1/\sqrt{t}$) can be clearly negligible, i.e.
$t^{2}V_{T(t)}(\phi(t))\;\rightarrow\;0$. Apart from very peculiar stiff
cases, all the inflationary potentials result to be negligible at very high temperatures.

Taking into account the above classical analysis, we may assume that, during
the Planck epoch, when the Universe performed its quantum evolution, the
potential of the scalar field plies no significant role; therefore, by
choosing the following expansion for the wave function%
\begin{equation}
\Psi(t,R_{c},\phi)=\int_{-\infty}^{\infty}\int_{-\infty}^{\infty}d\epsilon
dpC(\epsilon,p)\theta(\epsilon,pR_{c})exp\{\frac{i}{\hbar}(p\phi-\epsilon
t)\}\,,\label{a1}%
\end{equation}
(with $C(\epsilon,p)$ denoting generic coefficients), we get, from (\ref{z}),
the eigenvalues problem%
\begin{equation}
\left\{  \frac{l_{Pl}^{2}\hbar}{3\pi}\frac{d\quad}{dR_{c}}\frac{1}{R_{c}}%
\frac{d\quad}{dR_{c}}+\frac{p^{2}c}{4\pi^{2}}\frac{1}{R_{c}^{3}}+\frac{\mu
^{2}}{R_{c}}-\frac{3\pi\hbar}{4l_{Pl}^{2}}R_{c}\right\}  \theta=\frac
{\epsilon}{c}\theta\,.\label{a2}%
\end{equation}
with the boundary conditions $\theta(R_{c}=0)=0$ and $\theta(R_{c}%
\rightarrow\infty)=0$.\newline A solution to this equation reads in the form%
\begin{equation}
\theta\propto\sqrt{R_{c}}exp\left\{  -\frac{(R_{c}-R_{c(0)})^{2}}{4\alpha^{2}%
}\right\}  \,;\label{a3}%
\end{equation}
in order to be the above functional form a solution of equation (\ref{a2}), we
have to require the relations $p=\pm\sqrt{\pi\hbar/c}l_{Pl}$, $\alpha
=l_{Pl}/\sqrt{3\pi}$ and $\epsilon=-3\pi\hbar cR_{c(0)}/2l_{Pl}^{2}$.
Furthermore, since the ultra relativistic energy density is manifestly
positive, then, from the following expression for $\mu^{2}$%
\begin{equation}
\mu^{2}=\frac{l_{Pl}^{2}}{3\pi}\left(  \frac{1}{2\alpha^{2}}-\frac
{R_{c(0)}^{2}}{4\alpha^{4}}\right)  \,;\label{mu}%
\end{equation}
we find an important restriction on the continuous eigenvalue spectrum, i.e.
\begin{equation}
-\sqrt{3\pi/2}M_{Pl}c^{2}<\epsilon<\sqrt{3\pi/2}M_{pl}c^{2},
\end{equation}
being $M_{pl}$
the Planck mass, $M_{pl}=\hbar/cl_{Pl}$).\newline Thus, we get a
(non-normalizable) probability amplitude, for the stationary states, of the form%
\begin{equation}
P_{Stat}\propto\cos^{2}(\mid p\mid\phi)R_{c}exp\left\{  -\frac{(R_{c}%
-R_{c(0)})^{2}}{2\alpha^{2}}\right\}  \,.\label{a4}%
\end{equation}
The $\phi$-component of the wave function is not normalizable, because of the
potential field absence (we have to do with a situation analogous to that one
of a free non-relativistic particle admitting only two momentum eigenvalues) ,
but it is remarkable the existence, as effect of our revised quantization
approach, of stationary states for the radius of curvature; in the obtained
dynamics, we see that the notion of the cosmological singularity is replaced
by the more physical one of a peaked probability to find $R_{c}$ near zero.
The approximation of neglecting the potential term $V_{T}$ can be regarded as
confirmed \emph{a posteriori} by the small probability that the system
penetrates regions where $R_{c}$ is much greater than the Planck length and
the temperature is sufficiently small to be compared with the symmetry
breaking scale.

In order to construct the semiclassical limit of equation (\ref{a2}), we
separate $\theta$ into its modulus and phase, i.e. $\theta=\sqrt{\alpha}exp\{
i\beta/\hbar\}$; then we get the following two, real and complex, components
of equation (\ref{a2})%
\begin{equation}
-\frac{l_{Pl}^{2}}{3\pi\hbar} \frac{1}{R_{c}}\left(  \frac{d\beta}{dR_{c}}
\right)  ^{2} + \frac{p^{2}c}{4\pi^{2}}\frac{1}{R_{c}^{3}} + \frac{\mu^{2}%
}{R_{c}} - \frac{3\pi\hbar}{4l_{Pl}^{2}}R_{c} -\frac{\epsilon}{c} + \hbar^{2}
V_{Quantum} = 0 \, \quad\label{a5}%
\end{equation}%
\begin{equation}
\frac{1}{\sqrt{\alpha}} \frac{d\quad}{dR_{c}} \left(  \frac{\alpha}{R_{c}%
}\frac{d\beta}{dR_{c}} \right)  = 0 \Rightarrow\alpha\propto R_{c}%
/(d\beta/dR_{c}) \, , \label{a6}%
\end{equation}%
\begin{equation}
V_{Quantum}\propto\frac{1}{\sqrt{\alpha}}\frac{d\quad}{dR_{c}} \left(
\frac{1}{R_{c}}\frac{d\sqrt{\alpha}}{dR_{c}}\right)  \, . \label{a7}%
\end{equation}
In the limit $\hbar\rightarrow0$, when $V_{Quantum}$ becomes negligible, we
reobtain the Hamilton Jacobi equation describing the Universe classical
dynamics, but with an additional term corresponding to a non-relativistic
matter contribution, which, when $\epsilon$ is negative, acquires positive
energy density; to this respect, we remark how, on the quantum level, the
Universe is expected to approach the lowest, i.e. negative, energy
state.\newline We stress how, for sufficiently large $R_{c}$, if the
non-relativistic term dominates (the spatial curvature being yet negligible),
then we get $d\beta/dR_{c}\propto\sqrt{R_{c}}$ and therefore $R_{c}%
\rightarrow\infty\;\Rightarrow V_{Quantum}\sim1/(R_{c}^{3})\rightarrow0$; such
a behavior supports the idea that, when the Universe ``expands enough'' (i.e.
its volume fluctuating explores regions of high $R_{c}$ values), it can
approach a classical dynamics.

The analysis of this section answers the question about the cosmological
phenomenology implied by our approach and the issue goes toward the
appearance, in a synchronous reference, of a pressureless contribution to the
Universe energy density. In the next section, we make some estimations in
order to understand if such a new term (which is nothing more than the
classical limit of the total Universe quantum energy) may have something to do
with the observed dark matter component.

\subsection{Phenomenological Considerations}

\label{sec3 pc}

Indeed, by adding a term to the gravitational action, we may expect it appears
as a new kind of energy-momentum term; what makes our analysis a valuable one
is in the following points:\newline i) The kinematical action is an
embedding-like geometrical object, whose existence in quantum gravity, was
postulated in \cite{Mon2002} on the base of well-grounded statements and not
invented \emph{ad hoc}. Above we have shown that it can be interpreted, from a
classical point of view, as a non-relativistic dust fluid; a non-relativistic
energy density is also what appears from the quantum dynamics, when taking the
classical limit.\newline ii) All the accepted models of \emph{cold dark
matter} predict the existence of a very early (decoupled) zero-pressure
component, able, by this feature, to develop large scale structures (at the
present time even the \emph{heat dark matter} is expected to be
non-relativistic). Indeed, a non-baryonic component of this kind, is estimated
(either by the supernova data, either by the cosmic microwaves background
(detected) anisotropy) to be $\sim0.3$ of the actual Universe critical
density.\newline Since in equation (\ref{a5}) $\beta$ plays the role of the
(reduced) action function, we can write, by using Hamilton equations, the
following relation \footnote{the same result could be directly obtained by
applying the Hamilton-Jacobi method to the full action $\mathcal{S}%
=\beta(R_{c})+p\phi-\epsilon t$.}%
\begin{equation}
\frac{d\beta}{dR_{c}} = p_{R_{c}} = -\frac{3\pi\hbar}{2cl_{Pl}^{2}} R_{c}%
\frac{dR_{c}}{dt} \, . \label{pr}%
\end{equation}
Then, remembering that $H = (dR_{c}/dt)/R_{c}$ and $\bar{\Omega} - 1 =
c^{2}/H^{2}R^{2}_{c}$, we see how equation (\ref{a5}) takes the simple form
(with obvious notation for the different contributions) $\sum_{i} X_{i} = 1$,
being $X_{i}\equiv\bar{\Omega}_{i}/\bar{\Omega}$ ($i = p, \mu, dm, curv$);
thus, our dust fluid provides a component of the critical parameter
$\bar{\Omega}_{dm}$, given by%
\begin{equation}
\bar{\Omega}_{dm} = -\frac{4l_{Pl}^{2}c\epsilon}{3\pi\hbar H^{2}R_{c}^{3}}. \label{odm}%
\end{equation}
Such a formula is valid in general, independently of the other kinds of matter
present in the universe, and, therefore, provides a good tool to investigate
the role it could play in the actual cosmology; in this respect, we stress the
following three relevant points:\newline i) If we take for $\epsilon$ the
minimum value of the continuous spectrum obtained in the previous section,
within the framework of a ``pre-inflationary'' scenario, i.e. $\epsilon
\sim\mathcal{O}(-M_{Pl}c^{2})$, then we get%
\begin{equation}
\bar{\Omega}_{dm}=\mathcal{O}\left(  \frac{l_{Pl}c^{2}H^{-2}}{R_{c}^{3}%
}\right)  \sim\mathcal{O}(10^{-63})\,.\label{odm1}%
\end{equation}
ii) The value of $\epsilon$, required to have $\bar{\Omega}_{dm}%
=\mathcal{O}(1)$ (so that it could make account for the real dark matter
component, estimated about $0.3$ of the actual critical density), corresponds to%
\begin{equation}
\epsilon^{\ast}\sim\mathcal{O}\left(  \frac{\hbar cR_{c}^{3}}{l_{Pl}^{2}%
c^{2}H^{-2}}\right)  \sim\mathcal{O}(10^{82}GeV)\,;\label{odm2}%
\end{equation}
such a value corresponds to the present one of the total energy of the
Universe, whether it admits a closed space. A crucial point is that $\epsilon$
is a constant of the motion and therefore, since the Universe became a
classical one, it was characterized by such value $\epsilon^{\ast}$%
.\newline iii) In order to get an inflationary scenario, able to explain the
paradoxes of the Standard Cosmological Model, we need a sufficiently large
``e-folding'' which allows the size of an horizon, at the inflation beginning,
be now of the order of the actual Hubble radius; such a value corresponds, at
least, to about $60$, i.e. the ratio between the scale factors, respectively,
after and before the inflation, is around a factor $\mathcal{O}(10^{26})$.
This means that, if today $R_{c}\sim\mathcal{O}(10^{28}cm)$, then, taking into
account that the redshift of the end of the inflation is about $z\sim
\mathcal{O}(10^{24})$, we see that when the de-Sitter phase started its value
was $R_{c}\sim\mathcal{O}(10^{-22}cm)$. Thus, the total energy of the
Universe, when the dynamics became dominated by the ``vacuum energy'' at the
temperature $\sigma\sim\mathcal{O}(10^{15}GeV)$, is given by the expression%
\begin{equation}
\epsilon_{\Lambda}\sim\frac{\sigma^{4}R_{c}^{3}}{h^{3}c^{3}}\sim
\mathcal{O}(10^{36}GeV)\ll\epsilon^{\ast}\,;\label{odm3}%
\end{equation}
this result seems to indicate that, assuming the Universe underwent an
inflationary scenario, we get the contradictory issue about the impossibility
of a dominating ``vacuum energy''.

\section{Concluding Remarks}

We have presented a reformulation of the canonical
quantization of
geometrodynamics with respect
to a fixed reference frame; the main goal of our analysis is achieved by removing 
the fundamental shortcoming of the WDE stated at the point iii) in paragraph \ref{sec1 WDE}, 
i.e. now the 
quantization procedure takes place in a fixed reference frame and no 
ambiguity survives about 
the time-like character of the normal field; 
by other words, in this new approach it is possible to 
quantize the 3-geometry field on a fixed family of 
spatial hypersurfaces (corresponding to its 
evolution in the space-time), 
because this quantization scheme does not contradicts the 
strong assumption of a (3+1)-slicing 
of the 4-dimensional manifold. 

The main result
obtained, including the kinematical action
in the global dynamics, is the
characterization of an appropriate internal
physical clock. In our theory the
role of clock is played by the reference fluid,
comoving with the 3-hypersurfaces
and its presence is necessary to distinguish between
space-like and time-like geometrical objects
before the canonical quantization procedure.

The fluid shows its presence through a comoving
(non-positive defined) density of energy and momentum, which we
have characterized either from a classical
either from a quantum point of
view: classically it comes from
having introduced the kinematical action, but
its real nature must be investigated in the
classical limit of the eigenvalues equations.

It is worth noting that the considerations presented in paragraph \ref{sec3 pc} are against the idea that the here obtained $\bar{\Omega}_{dm}$
can make account for the dark matter, if inflation took place. The situation
is different if we take the picture of the Standard Cosmological Model
because, for instance, a classical estimation of the thermal bath energy at
the Planck epoch is about $\mathcal{O}((R_{c}/l_{Pl})^{3}M_{Pl}c^{2}%
)\sim\mathcal{O}(10^{112}GeV)$; thus, in absence of inflation, the value of
$\epsilon^{\ast}$ would have become important only in the later stage of the
Universe evolution and it could play today a relevant dynamical role.

Moreover to be applicable to a generic inhomogeneous gravitational
system, the theory here presented has to
be reduced, necessarily, to a formulation on a lattice;
recently some interesting proposal has appeared
to discretize a quantum constraint
\cite{DiBGamPul2002}, \cite{GamPul2002} and
they are of course relevant for the
discretization of the present theory. A more direct
approach can be obtained
applying the Regge calculus \cite{Reg1961}, \cite{Reg1997},
to the 3-geometries on the spatial hypersurfaces.

\appendix
\section{Multi-Time Approach}
\label{sec4 ap}

In this section we provide a schematic formulation of the so-called multi-time
approach and of its smeared Schr\"{o}dinger version, in view of a comparison
with the proposal of previous section.

The multi-time formalism is based on the idea that many gravitational degrees
of freedom appearing in the classical geometrodynamics have to be not
quantized because are not real physical ones; indeed we have to do with 10
$\times\infty^{3}$ variables, i.e. the values of the functions $N\,,N^{k}%
,\,h_{ij}$ in each point of the hypersurface $\Sigma^{3}$, but it is
well-known that the gravitational field possesses only $4\time\infty^{3}$
physical degrees of freedom in the phase space (in fact the gravitational
waves have, in each point of the space, only two independent polarizations and
satisfy second order equations).\newline The first step is therefore to
extract the real canonical variables by the transformation%
\begin{equation}
\left\{  h_{ij}\,\pi^{ij}\right\}  \,\rightarrow\,\left\{  \xi^{\mu}\,\pi
_{\mu}\right\}  \quad\left\{  H_{r}\,P^{r}\right\}  \quad\mu
=0,1,2,3\,r=1,2\,,\label{xax1}%
\end{equation}
where $H_{r},P^{r}$ are the four real degrees of freedom, while $\xi^{\mu
}\,\pi_{\mu}$ play the role of embedding variables.\newline In terms of this
new set of canonical variables, the gravity-``matter'' action (\ref{s}) rewrites%
\begin{equation}
S^{g\phi}=\int\limits_{\mathcal{M}^{4}}\left\{  \pi_{\mu}\partial_{t}\xi^{\mu
}+P^{r}\partial_{t}H_{r}+\pi_{\phi}\partial_{t}\phi-N(H^{g}+H^{\phi}%
)-N^{i}(H_{i}^{g}+H_{i}^{\phi})\right\}  d^{3}xdt,\label{xax2}%
\end{equation}
where $H^{g}=H^{g}({\xi}^{\mu},{\pi}_{\mu},H_{r},P^{r})$ and $H_{i}^{g}%
=H_{i}^{g}({\xi}^{\mu},{\pi}_{\mu},H_{r},P^{r})$.

Now we provide an ADM reduction of the dynamical problem by solving the
hamiltonian constraint for the momenta $\pi_{\mu}$%
\begin{equation}
{\pi}_{\mu}+h_{\mu}({\xi}^{\mu},H_{r},P^{r},\phi,\pi_{\phi})=0.\label{xax3}%
\end{equation}
Hence the above action takes the reduced form%
\begin{equation}
S^{g\phi}=\int_{\mathcal{M}^{4}}\left\{  P^{r}\partial_{t}H_{r}+\pi_{\phi
}\partial_{t}\phi-h_{\mu}\partial_{t}\xi^{\mu}\right\}  d^{3}xdt.\label{xax4}%
\end{equation}
Finally the lapse function and the shift vector are fixed by the hamiltonian
equations lost with the ADM reduction, as soon as, the functions $\partial
_{t}{\xi}^{\mu}$ are assigned. A choice of particular relevance is to set
$\partial_{t}{\xi}^{\mu}=\delta_{0}^{\mu}$ which leads to%
\begin{equation}
S^{g\phi}=\int_{\mathcal{M}^{4}}\left\{  P^{r}\partial_{t}H_{r}+\pi_{\phi
}\partial_{t}\phi-h_{0}\right\}  d^{3}xdt.\label{xax5}%
\end{equation}
The canonical quantization of the model follows by replacing all the Poisson
brackets with the corresponding commutators; if we assume that the states of
the quantum system are represented by a wave functional $\Psi= \Psi({\xi}%
^{\mu}, H_{r}, \phi)$, then the evolution is described by the equations%
\begin{equation}
i\hbar\frac{\delta\Psi}{\delta{\xi}^{\mu}}=\widehat{h}_{\mu}\Psi
\,,\label{xax6}%
\end{equation}
where $\widehat{h}_{\mu}$ are the operator version of the classical
hamiltonian densities.\newline In its smeared formulation the multi-time
approach reduces to the following Schr\"{o}dinger equation%
\begin{equation}
i\hbar\partial_{t}\Psi=\hat{\mathcal{h}}\Psi\,\quad\Psi=\Psi(t,H_{r}%
,\phi)\,.\label{xax7}%
\end{equation}
Here $\hat{\mathcal{h}}$ denote the quantum correspondence to the smeared hamiltonian%
\begin{equation}
\mathcal{h}=\int_{\mathcal{M}^{4}}\left\{  h_{\mu}\partial_{t}\xi^{\mu
}\right\}  d^{3}xdt.\label{xax8}%
\end{equation}
Now, observing that the first of equations (\ref{a1x}) can be rewritten as follows%
\begin{equation}
i\hbar\frac{\delta\Psi}{\delta y^{\mu}}=-n_{\mu}(\hat{H}^{g}+\hat{H}_{i}%
^{\phi})\Psi,\label{xax9}%
\end{equation}
it exists a correspondence between the above multi-time approach and our
proposal, viewed by identifying the formulas (\ref{azione totale}%
)-(\ref{xax5}), (\ref{xax9})-(\ref{xax6}) and (\ref{c1})-(\ref{xax7}). But the
following two key differences appear evident: i) the embedding variables
$y^{\mu}$ are added by hand, while the corresponding ones ${\xi}^{\mu}$ come
from non-physical degrees of freedom; ii) the hamiltonians $\mathcal{H}$ and
$\mathcal{h}$ (as well as their corresponding densities) describe very
different dynamical situations.

We show explicitly the parallel between these two approaches by their
implementation in a minisuperspace model: a Bianchi type IX Universe
containing a self-interacting scalar field. By using Misner variables
($\alpha$, $\beta_{+}$, $\beta_{-}$) \cite{MisThoWhe1973} the classical action
describing this system reads:%
\begin{align}
& S=\int\left\{  p_{\alpha}\dot{\alpha}+p_{\beta_{+}}\dot{\beta_{+}}%
+p_{\beta_{-}}\dot{\beta_{-}}+p_{\phi}\dot{\phi}-cNe^{-3\alpha}\times\right.
\nonumber\\
& \left.  \times-p_{\alpha}^{2}+p_{\beta_{+}}^{2}+p_{\beta_{-}}^{2}+p_{\phi
}^{2}+V(\alpha,\beta_{\pm},\phi)\right\}  dt,\,\quad c=const,\label{xax10}%
\end{align}
where $\overset{\cdot}{\left(  ....\right)  }\equiv\dfrac{d\left(
....\right)  }{dt}$and the precise form of the potential term $V$ is not
relevant for our discussion.\newline For this model, since the hamiltonian
density is independent of the spatial coordinates, then the multi-time
approach and its smeared Schr\"{o}dinger version overlap, the same being true
in our formalism.

In the spirit of our proposal the quantum dynamic of this model is described
by the equation%
\begin{equation}
i\hbar\partial_{t}\Psi=cNe^{-3\alpha}\hbar^{2}\left\{  \partial_{\alpha}%
^{2}-\partial_{\beta_{+}}^{2}-\partial_{\beta_{-}}^{2}-\partial_{\phi}%
^{2}+V\right\}  \Psi,\,\quad\Psi=\Psi(t,\alpha,\beta_{\pm},\phi),\label{xax11}%
\end{equation}
to which it should be added the restriction that the initial wave function
phase $\sigma_{0}=\sigma_{0}(\alpha,\beta_{\pm},\phi)$ satisfies the
Hamilton-Jacobi equation%
\begin{equation}
\left\{  -(\partial_{\alpha})^{2}+(\partial_{\beta_{+}})^{2}+(\partial
_{\beta_{-}})^{2}+(\partial_{\phi})^{2}\right\}  \sigma_{0}+V=0.\label{xax12}%
\end{equation}
In this scheme $N(t)$ is an arbitrary function of the label time to be
specified when fixing a reference.\newline To set up the multi-time approach
we have to preliminarily perform an ADM reduction of the dynamics
(\ref{xax10}). By solving the hamiltonian constraint obtained varying $N$, we
find the relation%
\begin{equation}
-p_{\alpha}\equiv h_{ADM}=\sqrt{p_{\beta_{+}}^{2}+p_{\beta_{-}}^{2}+p_{\phi
}^{2}+V}\,.\label{xax13}%
\end{equation}
Therefore action (\ref{xax10}) rewrites as%
\begin{equation}
S=\int\left\{  p_{\beta_{+}}\dot{\beta_{+}}+p_{\beta_{-}}\dot{\beta_{-}%
}+p_{\phi}\dot{\phi}-\dot{\alpha}h_{ADM}\right\}  dt\,.\label{xax14}%
\end{equation}
Thus we see how $\alpha$ plays the role of an embedding variable (indeed it is
related to the Universe volume), while $\beta_{\pm}$ are the real
gravitational degrees of freedom (they describe the Universe
anisotropy).\newline By one of the hamiltonian equation lost in the ADM
reduction (i.e. when varying $p_{\alpha}$ in (\ref{xax10})), we get%
\begin{equation}
\dot{\alpha}=-2cNe^{-3\alpha}p_{\alpha}=2cNe^{-3\alpha}h_{ADM}\,.\label{xax15}%
\end{equation}
Hence by setting $\dot{\alpha}=1$, we fix the lapse function as%
\begin{equation}
N=\frac{e^{3\alpha}}{2ch_{ADM}}\,.\label{xax16}%
\end{equation}
The quantum dynamics in the multi-time approach is summarized by the equation%
\begin{equation}
i\hbar\partial_{\alpha}\Psi=\sqrt{-\hbar^{2}(\partial_{\beta_{+}}^{2}%
+\partial_{\beta_{-}}^{2}+\partial_{\phi}^{2})+V}\Psi,\,\quad\Psi=\Psi
(\alpha,\beta_{\pm},\phi).\label{xax17}%
\end{equation}
We stress that in this multi-time approach the variable $\alpha$, i. e. the
volume of the Universe, behaves as a ``time``-coordinate and therefore the
quantum dynamics can not avoid the Universe reaches the cosmological
singularity ($\alpha\rightarrow-\infty$). On the other hand, in the formalism
we proposed, $\alpha$ is on the same footing of the other variables and are
admissible ``stationary states`` for which it is distributed in probabilistic
way.\newline This feature reflects a more general and fundamental difference
existing between the two approaches: the multi-time formalism violates the
geometrical nature of the gravitational field in view of real physical degrees
of freedom, while the proposed quantum dynamics implements this idea only up
to the lapse function and the shift vector, but preserves the geometrical
origin of the 3-metric field.

\end{document}